# Super-Arrhenius Dynamic Slowdown Revealed by Slow Variable Modulation in the Fragile Supercooled Liquid

Zhiye Tang[1,2], Shubham Kumar[1], and Shinji Saito[1,2,*]

[1]Institute for Molecular Science, Myodaiji, Okazaki, Aichi 444-8585 Japan

[2]The Graduate University for Advanced Studies (SOKENDAI), Myodaiji, Okazaki, Aichi 444-8585 Japan

*Corresponding author: shinji@ims.ac.jp

**Abstract**

The super-Arrhenius dynamic slowdown in fragile supercooled liquids remains one of the central unresolved questions in condensed matter physics. In this study, we analyze particle jump dynamics in a prototypical fragile glass-forming liquid, the Kob-Andersen Lennard-Jones (KALJ) model. Using the displacement of jumping particles as the reaction coordinate, we demonstrate the emergence of non-Poissonian dynamics as the temperature decreases. In the mildly supercooled regime, the outer region of the first coordination shell of a jumping particle exhibits a significant distribution shift during the jump motion. By comparing the survival probability with its slow-fluctuation limit using this distribution as a slow variable, we confirm that particles in this region modulate the jump dynamics, enhance the jump rate fluctuations, and thereby induce the dynamic slowdown as supercooling proceeds. As the temperature decreases, this behavior extends to the outer regions of the second coordination shell and beyond, intensifying the dynamic slowdown. This spatial growth of the slow variables responsible for dynamic disorder exhibits close correspondence with an increase in the static correlation length. These results provide a microscopic mechanism for the super-Arrhenius dynamic slowdown in the KALJ model.

## I. Introduction

A liquid is termed supercooled when it remains in the liquid state even after being below its melting point, owing to the suppression of crystallization. As the temperature decreases, molecular motions slow down and the relaxation time increases, yet this occurs without significant structural changes.[1] Eventually, when the relaxation time exceeds the timescale imposed by the cooling rate, the supercooled liquid falls out of equilibrium and begins to behave like a solid, and thus the glass transition occurs.

As a liquid is supercooled, its structural relaxation separates into two distinct processes: a fast β-relaxation associated with local motions within the cages formed by neighboring particles, and a slow α-relaxation resulting from particle diffusion beyond these cages.[2] The α-relaxation is typically non-exponential and often follows a stretched exponential form, which is generally attributed to dynamic heterogeneity.[3, 4] At deeper supercooling, particle jump motions between cages[5] become the more dominant mechanism driving α-relaxation.[6] Based on the temperature dependence of relaxation time, supercooled liquids are categorized as either strong or fragile.[7, 8] Strong liquids, such as silica melt,[9] exhibit Arrhenius behavior.[10] In contrast, fragile liquids display a super-Arrhenius temperature dependence, often interpreted in terms of a complex, hierarchical energy landscape where both spatial and temporal heterogeneity are more pronounced.[2, 3, 11, 12] Given the absence of any apparent structural changes during this process, typically characterized by two-point correlation functions, such as the radial distribution function and the static structure factor, the origin of the super-Arrhenius dynamic slowdown remains one of the central unresolved questions in condensed matter physics.[2, 13-15]

The dynamic slowdown accompanying supercooling has been extensively studied through both experimental and theoretical approaches, leading to a variety of theories addressing different aspects of the phenomenon. Dynamic facilitation, a concept developed from kinetically constrained models (KCMs), posits that structural relaxation occurs only when facilitated by neighboring active sites.[16-18] Mode coupling theory (MCT) offers a self-consistent treatment of density fluctuations and accurately describes the initial stages of the dynamic slowdown. However, it breaks down at deeper supercooling, as the predicted dynamic arrest is avoided by unaccounted activated processes.[19-21] The Adam-Gibbs theory attributes the slowdown to a reduction in configurational entropy.[22] This idea has been further developed in the random first-order transition (RFOT) theory, which introduces a static length scale characterizing cooperatively rearranging regions (CRRs), and predicts its growth as temperature decreases.[23, 24] Some efforts to measure static correlation lengths focused on local geometric indicators.[25-27] Alternatively, an order-agnostic approach, known as the point-to-set (PTS) static correlation length, was proposed. This framework establishes an inequality linking time and length scales, thereby offering a concrete method to evaluate the static correlation length within the RFOT perspective.[28-30] Interestingly, the static correlation length has been suggested to decouple from the dynamic correlation length, which characterizes dynamic heterogeneity.[31] It has also been shown to correlate with configurational entropy[32, 33] and to govern the growth of the α-relaxation time.[34] A systematic fragile-to-strong study using models with tunable parameters revealed the distinct temperature dependences of the static correlation lengths of fragile and strong liquids.[35] Nevertheless, the microscopic structural origins of this order-agnostic static length remain unclear.

Molecular dynamics (MD) simulations have been widely employed to study supercooled liquids, owing to their ability to resolve single-particle dynamics. Slow dynamics have been explored from the perspective of local geometric order, structural motifs, and geometric frustration.[12, 36-42] Analyses of the potential energy landscape have provided additional insight into the long-time diffusion and complex behavior of supercooled liquids.[43-46] Other studies have examined the relationship between local elasticity, soft vibrational modes, and the fragility of supercooled liquids.[47-50] Meanwhile, jump dynamics, which were identified both experimentally and computationally,[51-56] have been analyzed within the framework of the continuous-time random walk (CTRW) model, through which the long conjectured fast and slow phases in the supercooled liquid were confirmed.[57-59] The jump dynamics deviate from Poissonian statistics upon supercooling,[59, 60] exhibiting intermittent behavior attributed to dynamic disorder,[61] where environmental variables become as slow as the jump motion.[60, 62-64] Cooperative motions formed by consecutive jumps have been interpreted as manifestations of CRRs[65, 66] and dynamic facilitation[17]. An *a priori* theoretical *softness* has been introduced and shown to correlate strongly with the dynamics of supercooled liquids.[67, 68] It also correlates well with a machine-learning-based *softness*, which predicts the likelihood that a particle will undergo a jump in the near future.[69-71] More recently, advanced machine learning techniques have been applied to predict long-time dynamics from static structural information.[72-76] Despite these extensive efforts, the microscopic structural origin of the super-Arrhenius dynamic slowdown remains elusive.

This work addresses the question of why the dynamics of a supercooled liquid slow down more rapidly than predicted by the Arrhenius law in fragile liquids, despite the absence of significant structural changes. We investigate this phenomenon using the Kob-Andersen Lennard-Jones (KALJ) model,[77] an intensively studied prototypical fragile glass-forming liquid,[40, 42, 46, 78-91] by analyzing the particle jump motions. We find that particles in the outer region of the first coordination shell exhibit significant shifts in the distance distributions just below the onset of supercooling. We examine the survival probability of the cage state, which characterizes the jump dynamics, and its slow-fluctuation limits using the average distances to these particles as the slow variables. As a result, we demonstrate that these distances induce fluctuations in the jump rate, thereby modulating and slowing down jump motions. These dynamically relevant regions extend to the second and possibly more distant coordination shells as the temperature further decreases. As a result, the effective dimensionality of dynamics increases with decreasing temperature, leading to stronger dynamic disorder and a more pronounced dynamic slowdown. We also show a close correspondence between the spatial growth of slow variables and the PTS static correlation length.

## II. Method

### A. MD simulations

In this study, we performed MD simulations of the KALJ model.[77] The model is governed by the pair potential given by

$$v_{\alpha\beta}(r) = 4\epsilon_{\alpha\beta}\left[\left(\frac{\sigma_{\alpha\beta}}{r}\right)^{12} - \left(\frac{\sigma_{\alpha\beta}}{r}\right)^{6}\right], \quad (1)$$

where $\alpha$ and $\beta$ are either A or B. The interaction and radius are provided through $\epsilon_{AB}/\epsilon_{AA} = 1.5$, $\epsilon_{BB}/\epsilon_{AA} = 0.5$, $\sigma_{AB}/\sigma_{AA} = 0.8$, $\sigma_{BB}/\sigma_{AA} = 0.88$, respectively. The masses of the two species are equal, i.e., $m_A = m_B = 1$. The reduced units of the length, temperature, and time are given by $\sigma_{AA}$, $\epsilon_{AA}/k_B$, and $\sqrt{m_A\sigma_{AA}^2/\epsilon_{AA}}$, respectively. A time step of 0.001 was used for the time integration. The number density of the system was fixed at 1.2, and the numbers of particles A and B were $N_A = 800$ and $N_B = 200$, respectively. The interactions were truncated at $r = 2.5\sigma_{\alpha\beta}$. MD simulations were performed at temperatures of 0.982, 0.698, 0.550, 0.511, 0.476, and 0.455 in the microcanonical ensemble under periodic boundary conditions. All MD simulations were performed using LAMMPS,[92] and trajectories were saved at an interval of $\delta t = 0.01$.

### B. Definition of jump motions

Jump motion has been considered as the main elementary process underlying diffusion and structural relaxation at supercooling.[5] It has been identified and quantified using various methods, such as the Debye-Waller factor.[93] Here, we utilized the hop function proposed by Candelier et al,[94] defined as

$$h_i(t)=\sqrt{\left\langle\left(\boldsymbol{r}_i-\langle\boldsymbol{r}_i\rangle_{\Delta t_{\mathrm{B}}}\right)^2\right\rangle_{\Delta t_{\mathrm{A}}}\left\langle\left(\boldsymbol{r}_i-\langle\boldsymbol{r}_i\rangle_{\Delta t_{\mathrm{A}}}\right)^2\right\rangle_{\Delta t_{\mathrm{B}}}}, \tag{2}$$

where $\boldsymbol{r}_i$ is the coordinate of particle $i$, and $\langle\ldots\rangle_{\Delta t_{\mathrm{A}}}$ and $\langle\ldots\rangle_{\Delta t_{\mathrm{B}}}$ indicate the time averages over two adjacent time windows of the length $\Delta t/2$ centered at $t$, i.e., the intervals $[t - \Delta t/2, t]$ and $[t, t + \Delta t/2]$, respectively. $h_i(t)$ is the square root of the product of the distances between the coordinates in one time window and the mean position in the other. This function sharply increases when a jump occurs, i.e., when the particle resides in two different cages during time windows $\Delta t_{\mathrm{A}}$ and $\Delta t_{\mathrm{B}}$, respectively. Otherwise, it remains small. Thus, it provides a sensitive indicator of jump motion. Based on the physical interpretation, $\Delta t$ was chosen to be the time scale that minimizes the diffusivity,[93] defined as $\mathrm{dlog}\langle\Delta r^2(t)\rangle/\mathrm{dlog}t$, which is calculated from the mean squared displacement (MSD), $\langle\Delta r^2(t)\rangle$ (Fig. S1). This choice corresponds to the timescale of the motions within a cage, and it has been shown that the resulting jump dynamics remain qualitatively unchanged as long as $\Delta t$ lies between the short-time ballistic and long-time diffusive regimes.[60] The values of $\Delta t$ at various temperatures are provided in Table S2.

With $h_i(t)$ defined above, a jump event of particle $i$ was identified when $h_i(t)$ exceeded a threshold, $h^*$, such that the cage and jump states were defined as $h_i(t) < h^*$ and $h_i(t) \geq h^*$, respectively. In this study, $h^*$ was defined using the position where the slope of the cumulative probability distribution of the hop function, $F(h)$, changes from an initial fast decrease to a subsequent exponential decay (Fig. S2).[53, 95] The $h^*$ values were 0.18 and 0.35 for particles A and B, respectively, at all temperatures in this study.

### C. Jump dynamics

The residence time distribution, $\psi_{\mathrm{R}}(t)$, of the cage state, is obtained directly from the MD trajectories. Subsequently, the randomness parameter,[96, 97] $R$, was calculated via

$$R = \frac{\langle t^2\rangle-\langle t\rangle^2}{\langle t\rangle^2}, \tag{3}$$

where $\langle t^n\rangle = \int t^n\psi_{\mathrm{R}}(t)\mathrm{d}t$. For a Poissonian process, $\psi_{\mathrm{R}}(t)$ is described by an exponential function, i.e., $\psi_{\mathrm{R}}(t) = ke^{-kt}$, and thus $R = 1$.

The residence probability, $C_{\mathrm{R}}(t)$, and the survival probability, $C_{\mathrm{S}}(t)$, for the cage state can be obtained from $\psi_{\mathrm{R}}(t)$.[60, 98, 99]

$$C_{\mathrm{R}}(t) = \int_t^{\infty} \mathrm{d}t' \psi_{\mathrm{R}}(t'), \tag{4}$$

$$C_{\mathrm{S}}(t) = \frac{1}{\langle t \rangle} \int_t^{\infty} \mathrm{d}t' C_{\mathrm{R}}(t'). \tag{5}$$

$C_{\mathrm{R}}(t)$ represents the probability that a particle remains in the cage state for a duration of time $t$, whereas $C_{\mathrm{S}}(t)$ describes the probability that a particle that is currently in a cage state escapes from this cage state for the first time after a time $t$. When $R > 1$, $C_{\mathrm{S}}(t)$ decays more slowly than $C_{\mathrm{R}}(t)$, because the persistence time distribution has a longer tail than $\psi_{\mathrm{R}}(t)$.[17, 100] In this work, $C_{\mathrm{S}}(t)$ was calculated from $\psi_{\mathrm{R}}(t)$ using Eqs. (4) and (5), as obtaining the persistence time distribution directly would require much more extensive sampling of rare long-time jump events.[60, 101] Since the long-time tail of $\psi_{\mathrm{R}}(t)$ was adequately sampled, reliable statistics for $C_{\mathrm{S}}(t)$ were obtained.

The survival probability formally satisfies a stochastic rate equation,

$$\frac{\mathrm{d}C_{\mathrm{S}}(t)}{\mathrm{d}t} = -k(t) C_{\mathrm{S}}(t), \tag{6}$$

of which the formal solution is expressed as,[61]

$$C_{\mathrm{S}}(t) = \left\langle \exp\left( -\int_0^t k(t') \mathrm{d}t' \right) \right\rangle. \tag{7}$$

When $k(t)$ fluctuates much faster than the jump dynamics, the dynamics follow Poisson statistics, and the survival probability in this limit, $C_{\mathrm{fast}}(t)$, reduces to an exponential function,[60, 61, 98]

$$C_{\mathrm{fast}}(t) = \exp(-\langle k \rangle t). \tag{8}$$

where $\langle k \rangle$ is given by $1/\langle t \rangle$. On the other hand, when the fluctuation in $k(t)$ is much slower than the jump dynamics, the survival probability becomes a multi-exponential function weighted by the distribution of rates,[60, 61, 98]

$$C_{\mathrm{slow}}(t) = \langle \exp(-kt) \rangle_k. \tag{9}$$

Let $N_{\mathrm{T}}$ and $N^*$ be the total number of steps in the MD trajectories and the number of steps when $h$ reaches $h^*$, respectively. Then, the average rate in the fast-fluctuation limit is expressed by $\langle k \rangle = N^*/(N_{\mathrm{T}}\delta t)$. In the slow-fluctuation limit of $k(t)$, the jump dynamics depend on extra slow variables. Here, we assume the slow variable consists of $M$ states. Within the MD trajectories, the total number of steps in which the slow variable is in state $i$ is denoted by $N_i$, and the number of such steps when $h$ reaches $h^*$ is denoted by $N_i^*$. Subsequently, the rate of state $i$ is expressed by $k_i = N_i^*/(N_i\delta t)$. Note that $N_{\mathrm{T}} = \sum_i^M N_i$ and $N^* = \sum_i^M N_i^*$. The probability of state $i$ is defined by $p_i = N_i/N_{\mathrm{T}}$. With the $N_{\mathrm{i}}$, $N_{\mathrm{T}}$, and $N_{\mathrm{i}}^*$ values obtained from the MD trajectories, the survival probability in the slow-fluctuation limit is given by,[60, 61, 98]

$$C_{\mathrm{slow}}(t) = \sum_{i=1}^{M} p_i \exp(-k_i t) = \sum_{i=1}^{M} \frac{N_i}{N_{\mathrm{T}}} \exp\left(-\frac{N_i^*}{N_i \delta t} t\right). \quad (10)$$

Here, for a two-dimensional slow variable constructed from two constituent slow variables, the total number of states $M$ is given by the product of the numbers of states for each individual variable. Each state $i$ corresponds to a specific pair of values of the two slow variables, and the corresponding probability $p_{\mathrm{i}}$ represents their joint probability.

### D. Kullback-Leibler divergence

We quantified the difference between the probability distributions of a structural variable $x$ at equilibrium, $P_{\mathrm{eq}}(x)$, and *at* $h^*$, $P(x, h^*)$, by using Kullback-Leibler (KL) divergence,[102] defined as

$$D_{\mathrm{eq}} = \int \mathrm{d}x P_{\mathrm{eq}}(x) \log\left(P_{\mathrm{eq}}(x)/P\left(x, h^*\right)\right). \quad (11)$$

The $D_{\mathrm{eq}}$ values quantify the relative magnitudes of the structural changes associated with $x$ and thus allow identification of the slow variables. In addition, we also analyzed the difference between $P(x, h^*)$ and $P(x, h)$, where $P(x, h)$ denotes the probability distribution of $x$ at $h$, measuring the distribution change along $h$ towards $h^*$. It is defined as

$$D(h) = \int \mathrm{d}x P(x, h) \log\left(P(x, h)/P\left(x, h^*\right)\right). \quad (12)$$

$D(h)$, normalized by $D_{\mathrm{eq}}$, indicates how rapidly the distribution $P(x, h)$ evolves toward $h^*$, as a time-correlation function. For the evaluation of $P(x, h)$, $h$ was discretized with an interval of 0.005. Meanwhile, $P(x, h^*)$ was calculated by analyzing the $h_{\mathrm{i}}(t)$ trajectory of particle $i$ and selecting configurations at time steps where $h_{\mathrm{i}}(t)$ was immediately below $h^*$.

### E. Point-to-set correlation length

The PTS static correlation length can be evaluated by quantifying the confinement effect on a central particle when all particles located beyond a specified distance from the center are pinned. We calculated it according to the procedures established in previous works.[28, 103-107] Several pinning geometries have been employed in previous studies.[104] To appropriately evaluate the PTS correlation length, we used the cubic cavity geometry, which is similar in behavior to the spherical cavity.[104, 106, 108, 109] In this setup, all particles outside a cubic cavity of edge length $2d$, centered in the simulation box, were fixed. The value of $d$ was chosen between 1.3 and 3.9, depending on the temperature. The confinement effect on the particles in the center of the cavity was quantified using a grid-based overlap function[104] calculated as,

$$Q_{\alpha}(t, d) = \langle \sum_i n_{i,\alpha}(t) n_{i,\alpha}(0) \rangle / \langle \sum_i n_{i,\alpha}(0) \rangle, \tag{13}$$

where $n_{i,\alpha}(t)$ represents the occupation number of grid $i$ at time $t$, which takes one when the grid is occupied by $\alpha$ and zero otherwise. Here, $\alpha$ denotes particle A, B, or when omitted, either A or B, respectively. Because the overlap function depends on the grid size, $Q_\alpha(t,d\to\infty)$ decays to a value of $Q_\alpha(t\to\infty,d\to\infty) = \rho_\alpha v$, where $\rho_\alpha$ is the corresponding number density of $\alpha$, and $v$ is the volume of the grids. By subtracting $Q_\alpha(t\to\infty,d\to\infty)$ from $Q_\alpha(t\to\infty,d)$, we obtained the PTS correlation length $\xi_{\mathrm{pts},\alpha}$, which characterizes amorphous structural orders, using,

$$\tilde{Q}_{\alpha}(d) = Q_{\alpha}(t\to\infty, d) - Q_{\alpha}(t\to\infty, d\to\infty) = B_{\alpha}\exp\left[-\left(d\text{-}1/\xi_{\mathrm{pts},\alpha}\right)^{\eta_\alpha}\right], \tag{14}$$

where $B_\alpha$, $\eta_\alpha$, and $\xi_{\mathrm{pts},\alpha}$ are fitting parameters. To efficiently calculate the overlap function of the confined state, we performed simulations using the Monte Carlo (MC) parallel-tempering method,[107] except for a few cases with large $d$, where MD was performed (Tables S3 to S8). Conformations were saved every 100 production sweeps to evaluate the overlap function. A minimum of 43 initial configurations, each separated by at least 10 $\tau_\alpha$ were used to perform the MC or MD simulations under confinement. Here, $\tau_\alpha$ is the α-relaxation time defined as the time at which the self-intermediate scattering function, $F_S(k,t)$, decays to $e^{-1}$ (Fig. S3). A 5×5×5 grid with a spacing of 0.37 was used to calculate the overlap function.

## III. Results and discussion

### A. Structural changes accompanying the jump motion

First, we examined the structural changes accompanying the jump motion of particles. The jump motion has been considered a dominant contributor to diffusion and structural relaxation, particularly at low temperatures.[5] In this work, we analyzed the jump motion using the hop function, $h(t)$, which measures the displacement of a jumping particle. For consistency, we also used $h(t)$ at higher temperatures, where particle displacement occurs continuously. Representative time series of $h(t)$ within one α-relaxation time, $\tau_\alpha$, for particles A and B at $T = 0.455$ are shown in Figs. 1 (a) and 1 (e), respectively. This temperature lies deep in the supercooled regime, close to the MCT critical temperature, $T_c = 0.435$.[77] The $h(t)$ trajectories exhibit small, persistent fluctuations and occasional surges to large values, corresponding to local oscillations within a cage and large-amplitude displacements between cages, respectively.

The $h$-dependent radial distribution functions (RDFs) of jumping particles A and B, $g_A(r, h)$ and $g_B(r, h)$, at $T = 0.455$ were analyzed [Figs. 1 (b) and 1 (f) and Fig. S4]. In both $g_A(r, h)$ and $g_B(r, h)$, the coordination shell structure weakens as $h$ increases, showing lower peaks and higher minima, analogous to the effect of increasing temperature (Fig. S5). This behavior agrees with a previous study where fast and slow particles show less and more structured $g(r)$, respectively.[68] We defined the $h$-dependent coordination number, $CN(h)$, as the number of particles within the second minimum of $g_A(r)$ at ~1.4 and the first minimum of $g_B(r)$ at ~1.2 at all temperatures [Figs. 1 (b) and 1 (f)]. Here, the small peak around $r = 0.85$ in $g_A(r)$ originates mainly from the nearest neighboring particle B of particle A, and thus, the first coordination shell for particle A is defined at the second minimum. $CN_A(h)$ and $CN_B(h)$ decrease with increasing $h$ at all temperatures, primarily because the first peaks in the RDFs become lower [Figs. 1 (c) and 1 (g)]. At temperatures below 0.55, the decrease in $CN_B(h)$ begins with a rapid drop at $h < 0.05$, followed by a more gradual, nearly linear decline. The present results of the decrease in the neighboring particles along jump motions are consistent with the correlation between smaller $CN$ and larger jumping probability,[71] as well as the *softness* of particle rearrangement.[69]

Subsequently, we compared the equilibrium coordination numbers, $CN_{A,eq}$ and $CN_{B,eq}$, with $CN_A(h^*)$ and $CN_B(h^*)$, respectively [Figs. 1 (d) and 1 (h)]. $CN_{A,eq}$ remains nearly constant at 13.8 independent of the temperature, while $CN_{B,eq}$ decreases from 9.3 to 8.7 as the temperature decreases. The decrease in $CN_{B,eq}$ mainly arises from the contraction of the first minimum in $g_B(r)$ with decreasing temperature, which is attributed to changes in the $g_{BB}(r)$ component [Fig. S5 (e)]. These equilibrium values are consistent with the previous study.[110] Meanwhile, $CN_A(h^*)$ and $CN_B(h^*)$ decrease from 13.7 to 13.3 and from 9.1 to 8.1, respectively. For both particles A and B, the growing difference between $CN_{eq,\alpha}$ and $CN_\alpha(h^*)$ ($\alpha$ = A or B) at lower temperatures indicates a more pronounced decrease in neighboring particles along jump motion. These results are consistent with the previous study.[68] Such differences are less pronounced in the cases of $CN_{AA}$, $CN_{AB}$, $CN_{BA}$, or $CN_{BB}$ (Fig. S6), suggesting that $CN_{A,eq}$ and $CN_{B,eq}$, which account for both neighboring particles A and B, serve as better indices of the structural changes accompanying jump motion.

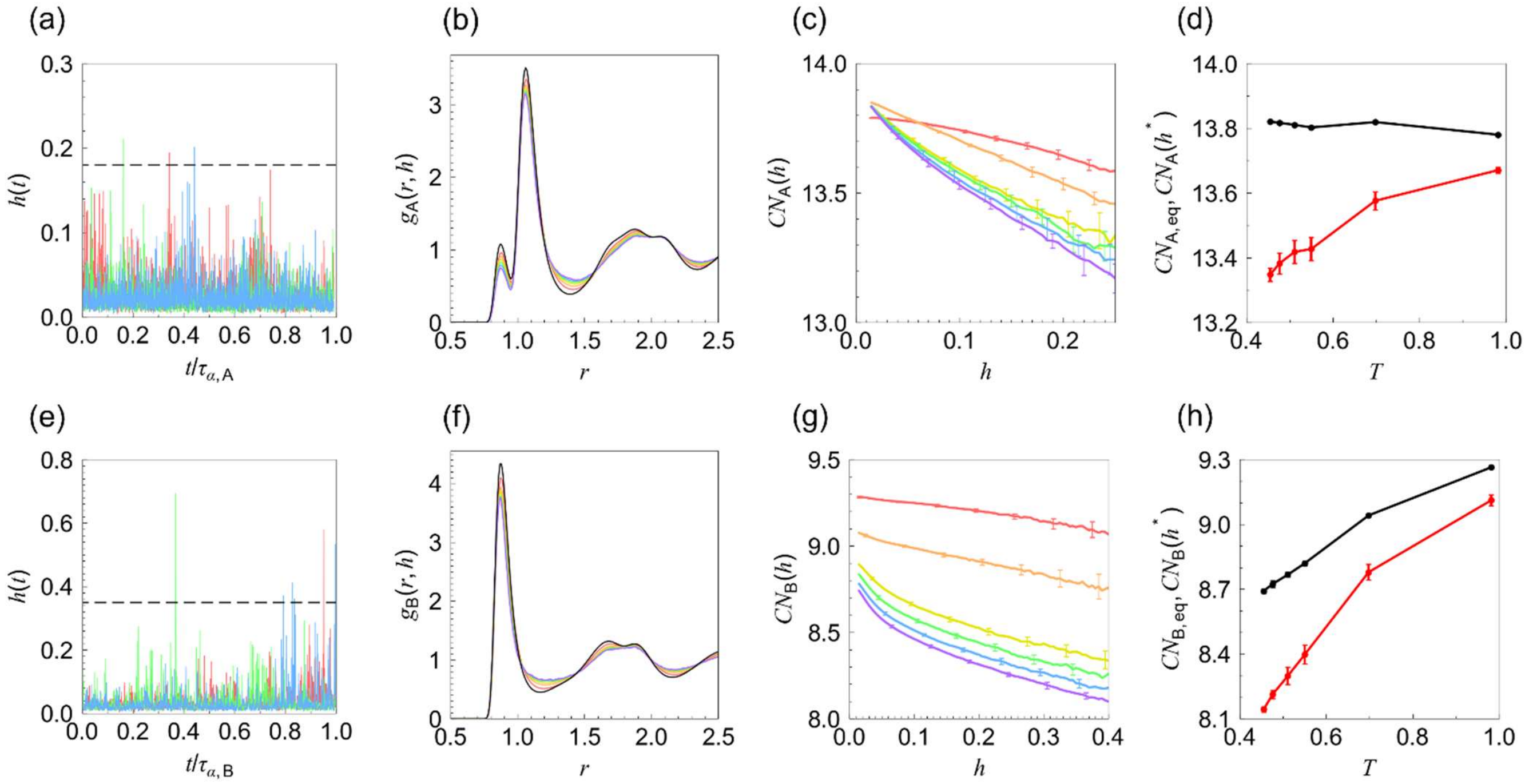

Figure 1. (a) Examples of $h(t)$ for three A particles. (b) $h$-dependent radial distribution function of particle A, $g_A(r,h)$, for $T$ = 0.455. (c) $h$-dependent coordination number of particle A, $CN_A(h)$. (d) Temperature-dependent coordination numbers, $CN_{A,eq}$ and $CN_A(h^*)$, of particle A at equilibrium (black) and at $h^*$ (red), respectively. (e)–(h) are the same as (a)–(d) but for particle B. In (b), colors represent $g_A(r,h)$ at $h$ = 0.040-0.045 (red), 0.075-0.080 (light red), 0.110-0.115 (yellow), 0.145-0.150 (green), 0.180-0.185 (blue), 0.215-0.220 (purple), together with the equilibrium result (black). In (c), colors represent $CN_A(h)$ at $T$ = 0.982 (red), 0.698 (orange), 0.550 (yellow), 0.511 (green), 0.476 (blue), and 0.455 (purple). In (f), colors represent $g_B(r,h)$ for $T$ = 0.455, at $h$ = 0.050-0.055 (red), 0.125-0.130 (light red), 0.200-0.205 (yellow), 0.275-0.280 (green), 0.350-0.355 (blue), 0.425-0.430 (purple), together with the equilibrium result (black). In (c), (d), (g), and (h), error bars represent the standard deviation evaluated across trajectory blocks.

We then investigated the correlation between the hop function of a jumping particle and those of the surrounding particles in its first four coordination shells. As the temperature decreases, the equilibrium values $h_{eq}$ of all particles decrease (Fig. S7), whereas the ratio $h_n/h_{eq}$ ($n = 1 – 4$) increases during jump motion (Fig. 2), due to stronger spatial correlations among neighboring particles. Interestingly, the $h_n/h_{eq}$ values for particle A are larger than those for particle B, indicating that the influence of a jumping particle A on its neighboring particles is greater than that of a jumping particle B (Fig. S8). Furthermore, the correlations remain noticeable up to the fourth coordination shell, particularly at lower temperatures, although they decrease from the inner to the outer shell at each temperature (Fig. 2). These results are consistent with a previous finding that the spatial correlation of $h$ decays exponentially.[71] The spatial correlations found in this study are consistent with the previous results showing that long-time structural relaxation is more strongly coupled to order parameters averaged over neighboring particles than to single-particle order parameters, with the relevant averaging length extending over approximately four coordination shells.[27, 91, 111] These features are reminiscent of the dynamic facilitation volume, in which the facilitation among neighboring particles leads to correlated motions in supercooled liquids.[112, 113]

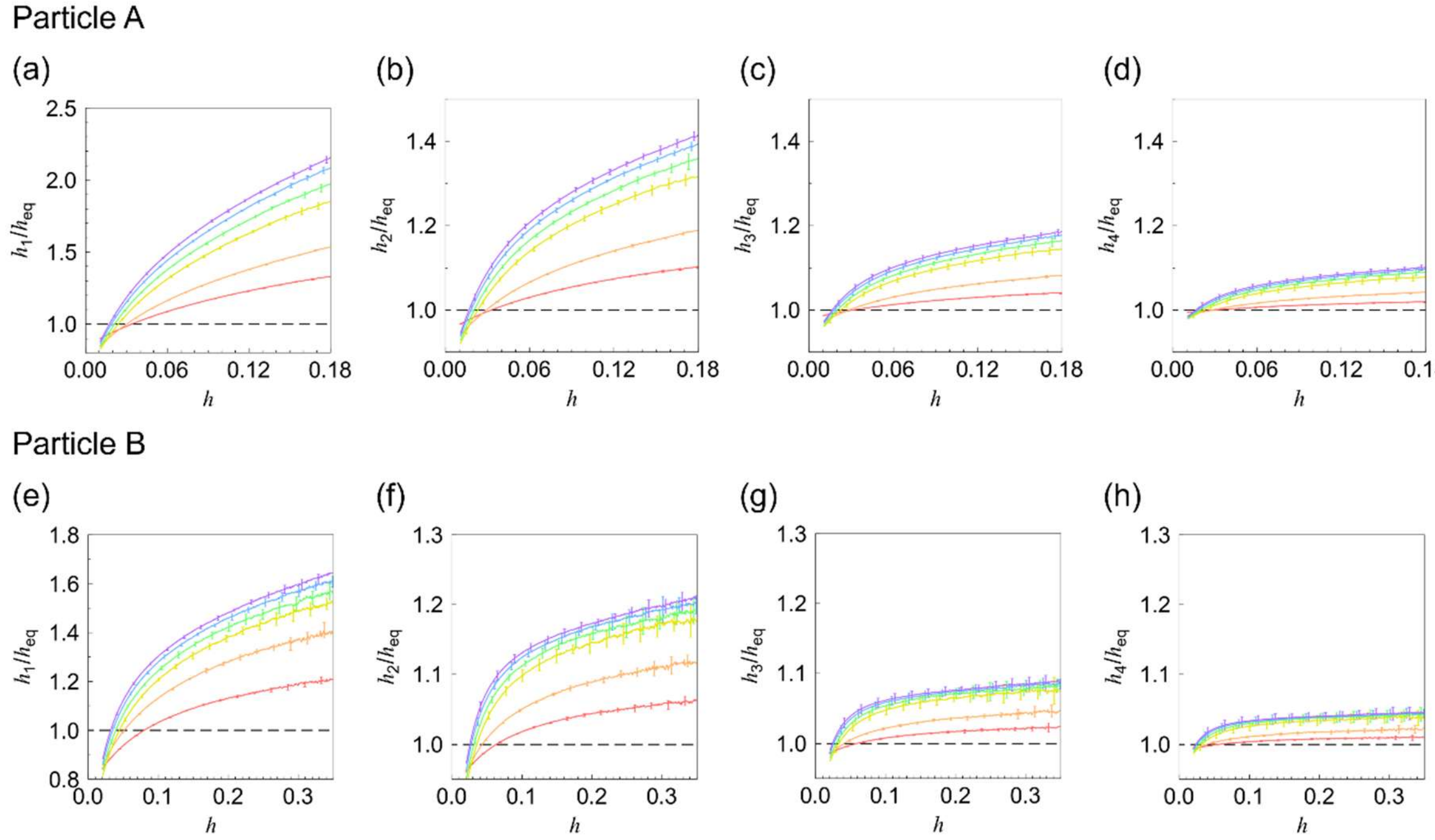

Figure 2. $h_n/h_{\mathrm{eq}}$ ($n = 1 - 4$) averaged over neighboring particles in the (a) first, (b) second, (c) third, and (d) fourth coordination shells of particle A, and in the (e) first, (f) second, (g) third, and (h) fourth coordination shells of particle B. The colors represent $T = 0.982$ (red), 0.698 (orange), 0.550 (yellow), 0.511 (green), 0.476 (blue), and 0.455 (purple). Error bars represent the standard deviation evaluated across trajectory blocks.

**B. Changes in jump dynamics with decreasing temperature**

We examined the temperature dependence of jump dynamics by analyzing the distribution of the residence time in the cage state, $\psi_{\mathrm{R}}(t)$. $\psi_{\mathrm{R}}(t)$ exhibits a strong temperature dependence. At $T = 0.982$, which lies above the onset of supercooling at $T = 0.8$,[114, 115] $\psi_{\mathrm{R}}(t)$ follows an exponential form [red curves in Figs. 3 (a) and 3 (b)], indicating that the jump dynamics follow Poisson statistics. As the temperature decreases, the shape of $\psi_{\mathrm{R}}(t)$ evolves into a stretched exponential form[116] with an elongated tail [e.g., purple curves in Figs. 3 (a) and 3 (b)], in agreement with previous findings.[59] This result indicates a deviation from Poisson statistics, revealing that the jump dynamics involve multiple timescales rather than a single one.[117] To further quantify this deviation, we calculated the randomness parameter,[96] $R$, for particles A and B, respectively [Fig. 3 (c)]. The $R$ values for particles A and B begin to deviate from unity at $T \approx 0.6$ and sharply rise to 2.83 and 3.87, respectively, at $T = 0.455$. This behavior signifies a transition from a Poissonian to a renewal process, with increasingly intermittent dynamics.[118]

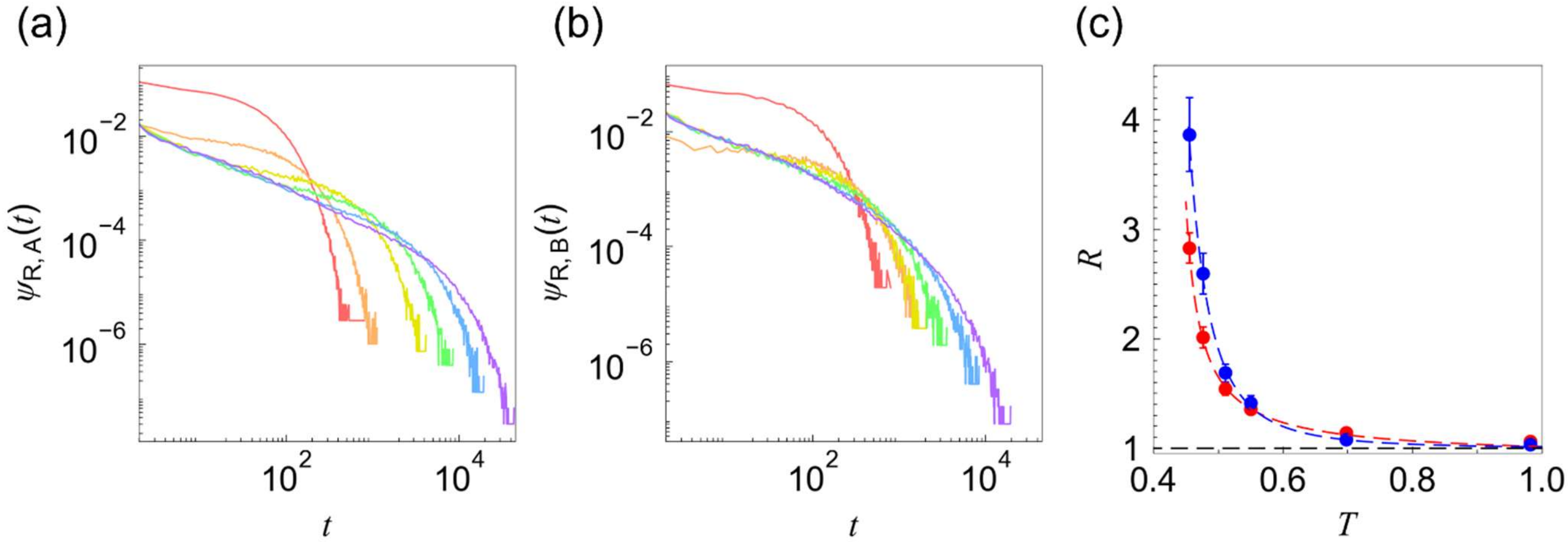

Figure 3. (a) and (b) Residence time distributions for particles A and B, respectively. (c) Randomness parameter, $R$, for particles A (red) and B (blue). In (a) and (b), the color codes are the same as those in Fig. 2. In panel (c), error bars represent the standard deviation evaluated across trajectory blocks.

We subsequently investigated the jump dynamics using the survival probability of the cage state, $C_S(t)$, and its fast-fluctuation limit, $C_{fast}(t)$. At $T = 0.982$, $C_S(t)$ coincides with $C_{fast}(t)$, indicating that the jump rate is well described by its time average $\langle k \rangle$ in the fast-fluctuation limit [Fig. 4 (a)]. However, as the temperature decreases below 0.550, $C_S(t)$ becomes progressively slower than $C_{fast}(t)$ and adopts a stretched exponential form.

To characterize the rate fluctuations, $C_S(t)$ was fitted with a stretched exponential function, $C_S(t) = \exp[-(kt)^{\beta}]$ [Fig. 5 (a)]. Above $T = 0.7$, the exponent $\beta \approx 1$ indicates that $h$ exhibits the slowest dynamics among all variables, thereby suggesting that $h$ serves as an effective reaction coordinate for the jump dynamics. As the temperature decreases below $T = 0.550$, $\beta$ deviates noticeably from unity, and this trend intensifies [Fig. 5 (a)], indicating increasingly substantial fluctuations in $k(t)$ caused by the presence of slow variables other than $h$, i.e., the progressive development of dynamic disorder.[61]

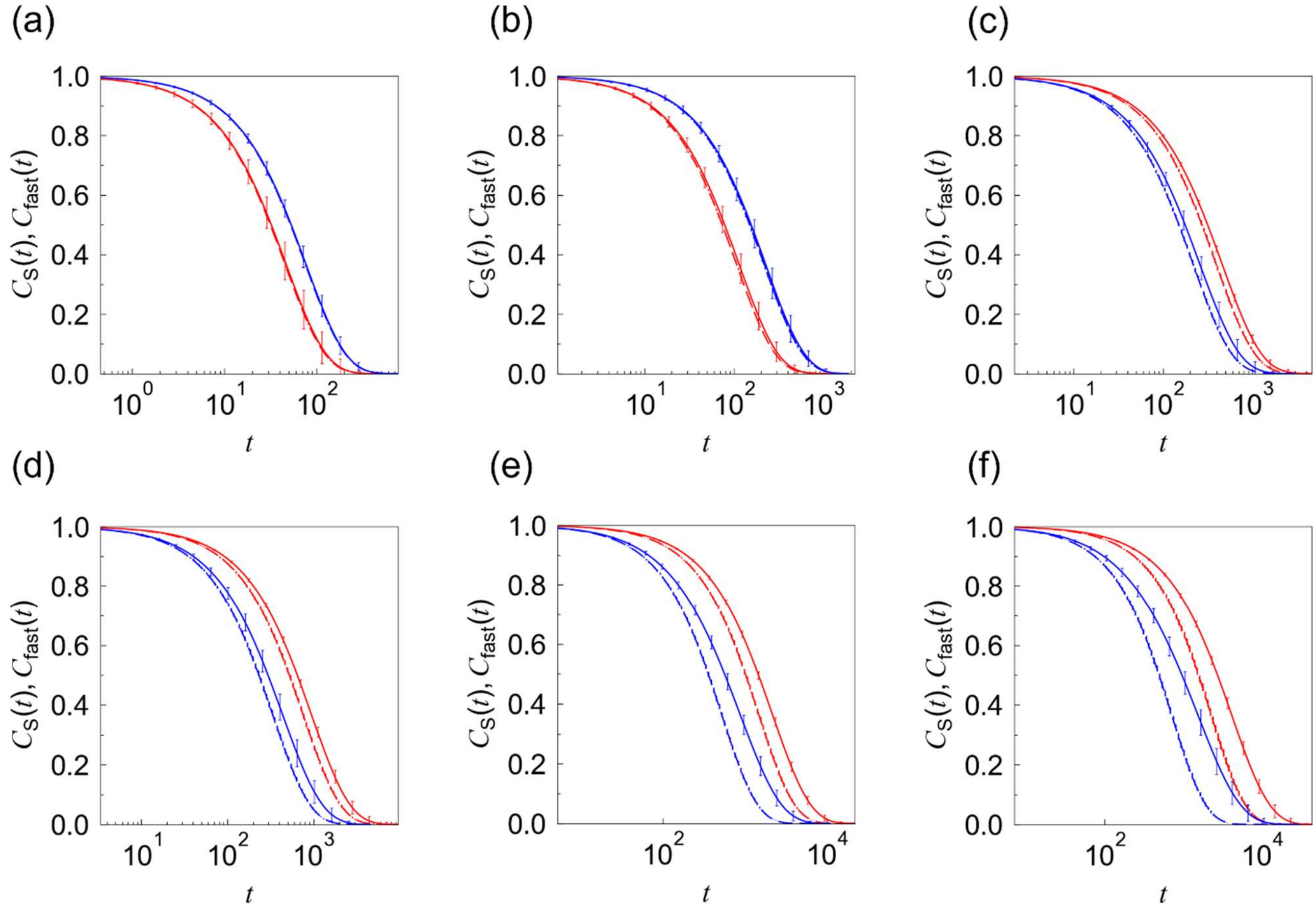

Figure 4. Survival probability, $C_S(t)$ (solid), and its fast-fluctuation limit, $C_{fast}(t)$ (dashed) for particles A (red) and B (blue) at $T = 0.982$ (a), 0.698 (b), 0.550 (c), 0.511 (d), 0.476 (e), and 0.455 (f). Error bars represent the standard deviation evaluated across trajectory blocks.

To quantify the fluctuations in $k(t)$, we also analyzed the rate distribution $P(k)$, obtained from the inverse Laplace transform[119] (iLT) of $C_S(t) = \int_0^\infty P(k)\exp(-kt)dk$ [Figs. 5 (b) and 5 (c)]. At $T = 0.982$ and 0.698, $P(k)$ exhibits a single peak centered at $\langle k \rangle$. However, when the temperature decreases to $T \leq 0.550$ (corresponding to $\beta < 1$), multiple peaks emerge in both $P_A(k)$ and $P_B(k)$. The appearance of these peaks suggests that motions spanning a broad range of timescales become involved in the jump dynamics at these temperatures. These results agree with the larger fluctuations in $k(t)$ [Fig. 5 (a)].

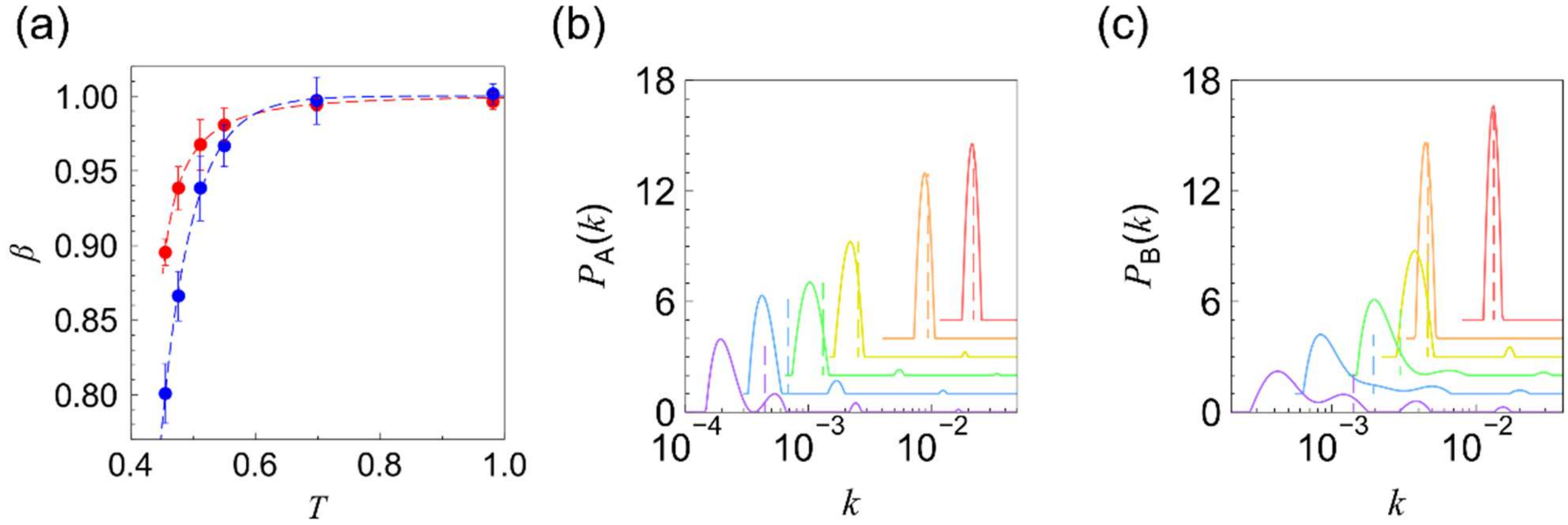

Figure 5. (a) Exponent $\beta$ values obtained by fitting $C_S(t)$ to the stretched-exponential form, $C_S(t) = \exp[-(kt)^{\beta}]$, for particles A (red) and B (blue). (b) and (c) Distributions of the rate $k$ for particles A and B, respectively, obtained from the inverse Laplace transform of $C_S(t)$. In (a), error bars represent the standard deviation evaluated across trajectory blocks. In (b) and (c), the color codes are the same as those in Fig. 2. Vertical dashed lines indicate the average rate $\langle k \rangle$. Curves are shifted upward by one unit for each temperature.

### C. Structural origins of dynamic disorder

To identify the slow variables that compete with $h$ and cause fluctuations in $k(t)$, we analyzed the structural changes associated with jump events by examining $D_{eq}$ [Eq. (11)], comparing the equilibrium distribution of $r_n$, $P_{eq}(r_n)$, with the corresponding distribution at $h^*$, $P(r_n,h^*)$, for neighboring particles of particles A and B [Figs. 6 (a)–6 (d) and S9].[60] Here, $r_n$ denotes the distance between a jumping particle and its $n$-th nearest neighbor. As described above, $CN_{A,eq}$ and $CN_{B,eq}$ are more appropriate indices of structural changes than coordination numbers defined by distinguishing particle species such as $CN_{AA}$. At $T$ = 0.455, for particle A, two pronounced groups of $D_{eq}$ values indicating greater displacements appear in the outer regions of the first and second coordination shells, which consist of the 9th to 14th and 31st to 60th nearest neighbors, respectively [Fig. 6 (a)]. The outer half of the first coordination shell dominates the increase in $CN$ change associated with jump motion at lower temperatures in Fig. 1 (d). In contrast, the inner halves of the coordination shells exhibit smaller $D_{eq}$ values. The magnitude of $D_{eq}$ increases with decreasing temperature, indicating that larger structural changes are required for particles to perform a jump; for example, $D_{eq}$ is approximately twice as large at $T$ = 0.455 compared with $T$ = 0.982 [Figs. 6 (a) and 6 (b)]. A similar behavior is observed for particle B: pronounced peaks in $D_{eq}$ appear in the outer halves of the first and second coordination shells. Because particle B is smaller and thus has fewer neighboring particles, the particles contributing to these regions correspond to the sixth to ninth and 25th to 46th nearest neighbors [Figs. 6 (c), 6 (d), and S9 (g)–S9 (l)].

In the KALJ model, the number of particles involved in jump motions that contribute to relatively large $D_{eq}$ values is much greater than the corresponding number of molecules in supercooled water and silica melt, [60, 101] whereas the $D_{eq}$ value itself is smaller. This contrast reflects the fact that the denser packing in the KALJ model necessitates the cooperative participation of a large number of particles in a single jump event. To characterize the cooperativity associated with jump motions, we analyze the jump dynamics using collective variables, the average distances to subsets of particles exhibiting large $D_{eq}$ values. Specifically, we define $\bar{r}_{n\text{-}m}$ as the average distance between a jumping particle and its $n$-th to $m$-th nearest neighbors.

At $T = 0.455$, for particle A, the most important collective variable, the average distance to the 9th to13th nearest neighbors, denoted as $\bar{r}_{9\text{-}13}$, exhibits a substantially larger $D_{eq}$ value of 0.595, compared with 0.494 of $r_{12}$. It also exceeds the value of 0.570 for $\bar{r}_{9\text{-}14}$ consisting of the 9th to 14th nearest neighbors (Fig. S10), because the 14th neighbor lies at the boundary and shows a non-aligned probability shift that reduces the $D_{eq}$ value of $\bar{r}_{9\text{-}14}$[Figs. S11 (a)–S11 (f)]. Therefore, we identified $\bar{r}_{9\text{-}13}$ as the collective variable for the first coordination shell, *CV1* [Fig. 6 (e)]. Similarly, we defined $\bar{r}_{31\text{-}60}$ consisting of the 31st to 60th nearest neighbors, identified above with $D_{eq} = 0.309$, as *CV2* for the second coordination shell of particle A [Fig. 6 (f)]. Because *CV2* involves the average over many neighboring particles, small variations in the particle range do not quantitatively affect the resulting $D_{eq}$ values. For particle B, $\bar{r}_{6\text{-}8}$ (6th to 8th) with $D_{eq} = 0.698$ was defined as *CV1* [Figs. S10 and S11 (g)–S11 (j)], and $\bar{r}_{25\text{-}47}$ (25th to 47th) with $D_{eq} = 0.264$ was defined as *CV2*. *CV1* and *CV2*, together with their distance distributions at the other temperatures, are shown in Figs. S12 and S13.

To quantify changes in distributions of collective variables along jump motion, we evaluated $D(h)$. To facilitate a direct comparison of the $h$ dependence between *CV1* and *CV2*, we used the normalized $D(h)$ by $D_{eq}$ [Figs. 6 (i)-6 (l), S14, and S15]. As a result, the decay of $D(h)/D_{eq}$ in *CV1* is slower than that in *CV2*, and this difference becomes more pronounced as the temperature decreases [Figs. 6 (j), 6 (l), and S15]. This indicates a hierarchical pathway for jumping dynamics, in which rearrangement in the second shell precedes that in the first shell and subsequently triggers jump events. Similar precursor rearrangements have been found in previous studies of supercooled water and silica melt,[60, 101] and are consistent with the dynamic facilitation picture (Fig. S15).[112]

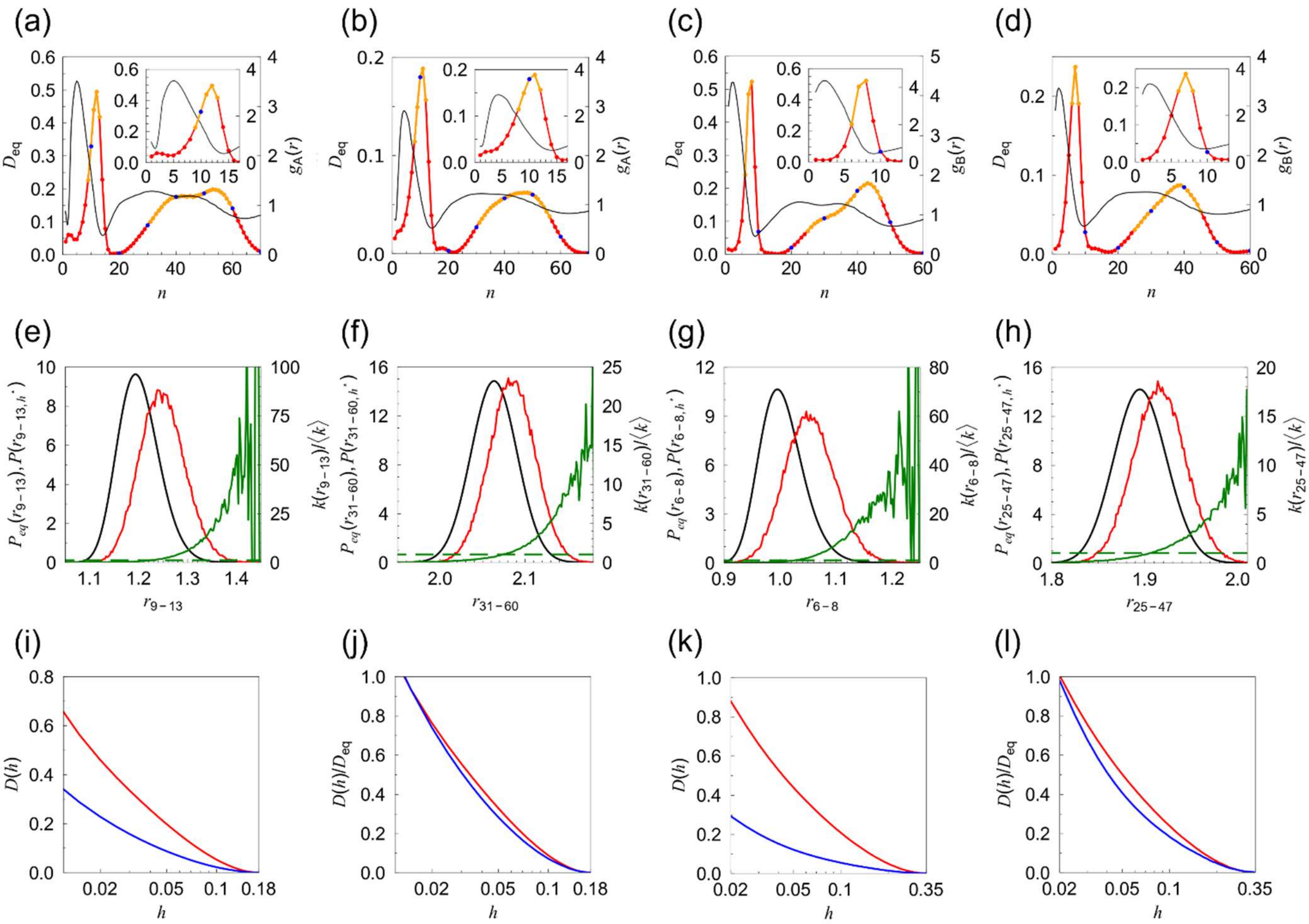

Figure 6. (a)-(d) $D_{\mathrm{eq}}$ (red) and $g_{\mathrm{A}}(f(\langle r_n \rangle))$ (black) for neighboring particle $n$, plotted against $n$ for particle A at $T = 0.455$ (a) and $T = 0.982$ (b), and for particle B at $T = 0.455$ (c) and $T = 0.982$ (d). (e)-(h) Equilibrium distance distributions $P_{\mathrm{eq}}(x)$ (black) and the distributions at $h^*$, $P(x,h^*)$ (red), together with the relative rate, $k(x)/\langle k \rangle$ (green, right axis) at $T = 0.455$, for collective variables $x$ selected as $\bar{r}_{9\text{-}13}$ (e) and $\bar{r}_{31\text{-}60}$ (f) for particle A, and $\bar{r}_{6\text{-}8}$ (g) and $\bar{r}_{25\text{-}47}$ (h) for particle B, respectively. (i)-(l) $D(h)$ and $D(h)/D_{\mathrm{eq}}$ for particles A in (i) and (j) and B in (k) and (l). In (a)-(d), $\langle r_n \rangle = \int r_n P_{\mathrm{eq}}(r_n) \mathrm{d}r_n$, and $f(\langle r_n \rangle)$ is a piecewise linear function that maps $\langle r_n \rangle$ to $n$. Blue dots mark every tenth particle, and orange dots highlight the particles in *CV1* and *CV2*; inset panels show enlargements of the first peaks. In (i)-(l), red curves correspond to *CV1* and blue curves correspond to *CV2*.

Subsequently, we evaluated the survival probabilities in the slow-fluctuation limit $C_{\mathrm{slow},1}(t)$, $C_{\mathrm{slow},2}(t)$, and $C_{\mathrm{slow},1\text{-}2}(t)$ by assuming that *CV1*, *CV2*, and the composite variable *CV1-2* constructed from *CV1* and *CV2* are slow variables, respectively (Figs. 7 and 8). We first analyzed the survival probabilities for particle A. At $T = 0.982$, where the jump dynamics are Poissonian, the reaction coordinate $h$ alone appropriately describes the jump dynamics. At this temperature, variables other than $h$, including *CV1*, fluctuate much faster than the jump dynamics governed by $h$, and are therefore averaged out as fast degrees of freedom along the jump process. Consequently, as shown in Fig. 7 (a), all the survival probabilities in the slow-fluctuation limit decay more slowly than in the actual survival probability $C_{\mathrm{S}}(t)$. At $T = 0.698$, $C_{\mathrm{S}}(t)$ becomes slightly slower than $C_{\mathrm{fast}}(t)$, indicating the involvement of at least one slow variable [Fig. 7 (b)]. We therefore calculated $C_{\mathrm{slow,1}}(t)$ by adopting *CV1* as a slow variable, and found that $C_{\mathrm{slow,1}}(t)$ decays even more slowly than $C_{\mathrm{S}}(t)$ over the entire timescale. Thus, while *CV1* serves as a primary slow variable that effectively accounts for $C_{\mathrm{S}}(t)$, the fluctuations of *CV1* that contribute to the actual jump dynamics are not fully in the slow-fluctuation limit, i.e., their characteristic timescale remains faster than that of the time evaluation of $h$. At $T = 0.511$, $C_{\mathrm{slow,1}}(t)$ decays more slowly at long times but slightly faster at short times than $C_{\mathrm{S}}(t)$. This indicates that *CV1* alone is insufficient to describe $C_{\mathrm{S}}(t)$ [Fig. 7 (d)]. We therefore introduce an additional slow variable *CV2*, which alone decays faster than $C_{\mathrm{S}}(t)$, and find that $C_{\mathrm{slow,1\text{-}2}}(t)$ mitigates the faster decay at short timescales in $C_{\mathrm{slow,1}}(t)$ compared with $C_{\mathrm{S}}(t)$. These results indicate that at this temperature, at least two slow variables, i.e., *CV1* and *CV2*, contribute to the fluctuations in the jump dynamics. As the temperature further decreases to $T = 0.455$, however, even $C_{\mathrm{slow,1\text{-}2}}(t)$ decays faster than $C_{\mathrm{S}}(t)$ at $t < 2\times10^3$, suggesting that additional slow variables are required to fully characterize the jump dynamics at lower temperatures [Fig. 7 (f)]. For particle B as well, incorporating the slow variables *CV1* and *CV2* becomes essential to properly describe the jump dynamics as the temperature decreases (Fig. 8). These results indicate that as the temperature decreases, the cooperativity in jump dynamics grows spatially, necessitating the introduction of slow variables located further from the jumping particle. Similar spatial growth has been found in water and silica melt,[60, 101] with decreasing temperature, and is consistent with the hierarchical picture of jump dynamics shown above.

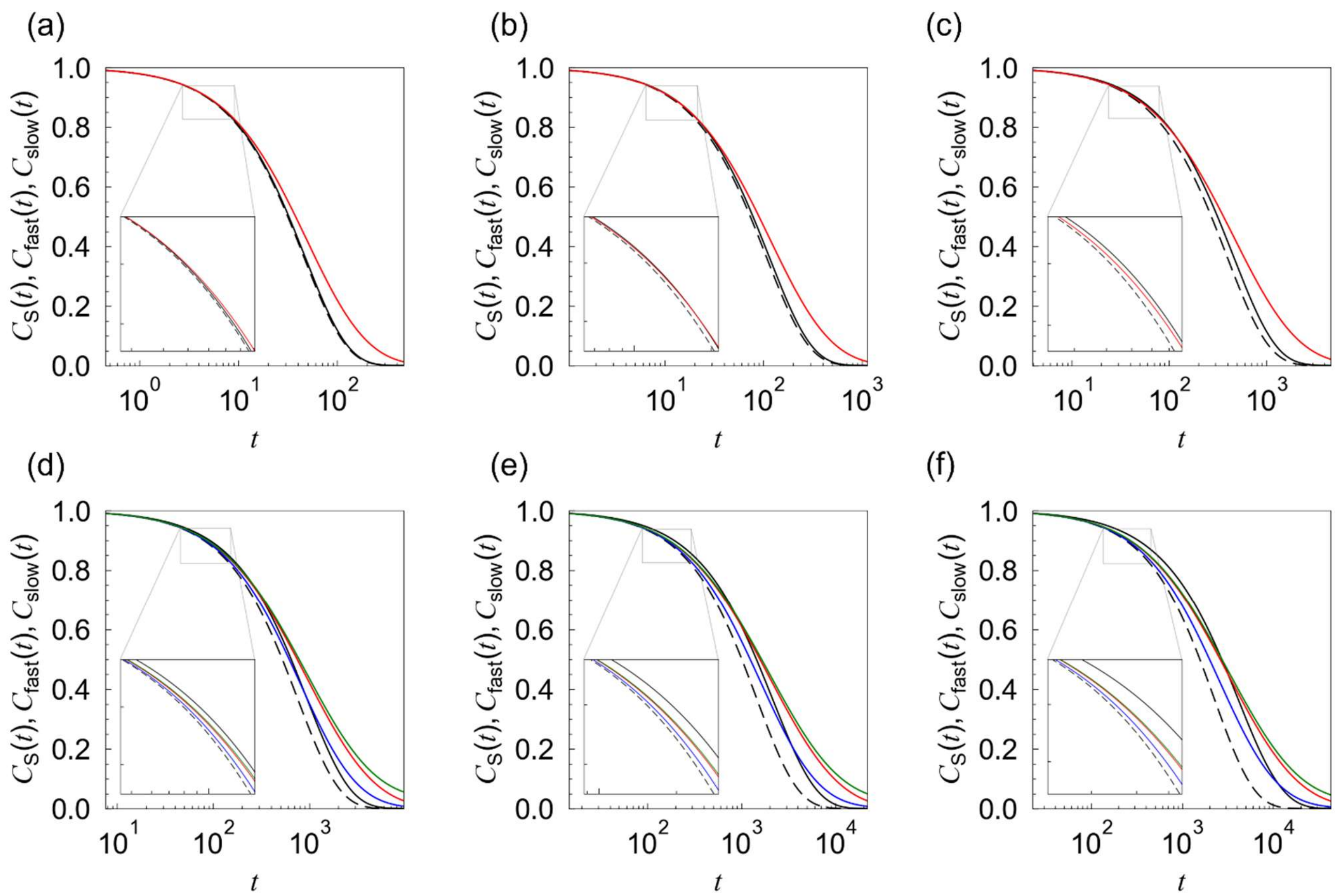

Figure 7. Survival probability, $C_S(t)$ (black solid), its fast-fluctuation limit, $C_{fast}(t)$ (black dashed), and slow-fluctuation limits, $C_{slow,1}(t)$ (red), $C_{slow,2}(t)$ (blue), and $C_{slow,1\text{-}2}(t)$ (green) for particle A at $T$ = 0.982 (a), 0.698 (b), 0.550 (c), 0.511 (d), 0.476 (e), and 0.455 (f).

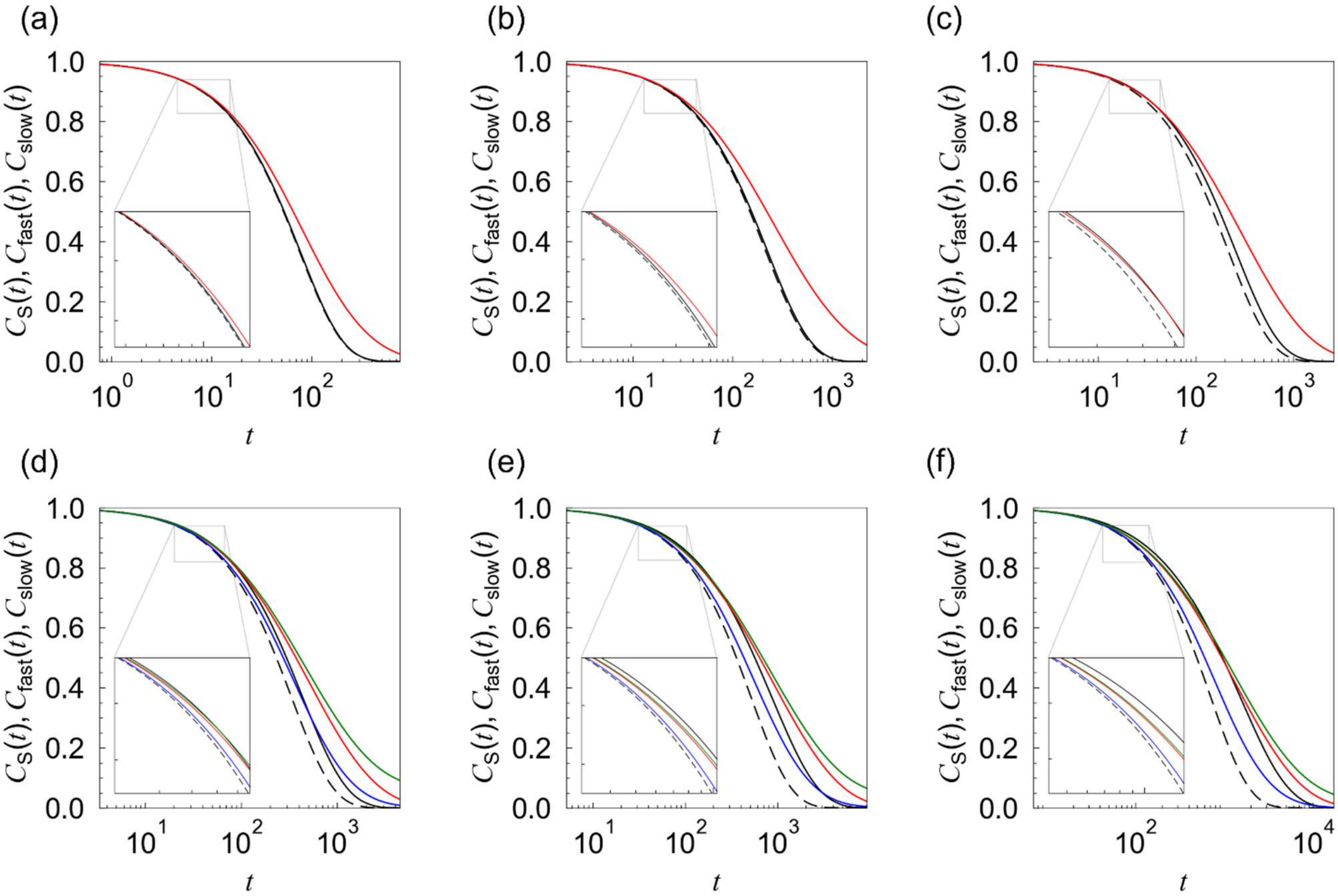

Figure 8. Survival probability, $C_S(t)$ (black solid), its fast-fluctuation limit, $C_{fast}(t)$ (black dashed), and slow-fluctuation limits, $C_{slow,1}(t)$ (red), $C_{slow,2}(t)$ (blue), and $C_{slow,1\text{-}2}(t)$ (green) for particle B at $T$ = 0.982 (a), 0.698 (b), 0.550 (c), 0.511 (d), 0.476 (e), and 0.455 (f).

Furthermore, we compared the rate distributions $P(k)$ obtained from $C_S(t)$ at each temperature with those from the survival probability in the slow limit (Fig. 9). At higher temperatures ($T > 0.550$), $C_S(t)$ is well approximated by $\langle k \rangle$ in the fast limit. In this regime, the rate distribution $P_{slow,1}(k)$ (red line), obtained from $C_{slow,1}(t)$, is much broader than $P(k)$. As noted above, upon lowering the temperature below 0.550, multiple peaks emerge in $P(k)$ for both particles A and B. At these temperatures, $P_{slow,1}(k)$ remains broad, spreading to both sides of the main peaks of $P(k)$, i.e., toward smaller $k$ and larger $k$. Consequently, $C_{slow,1}(t)$ decays faster at short times but more slowly at long times than $C_S(t)$ [Figs. 7(d) and 8(d)]. The rate distribution $P_{slow,2}(k)$ (blue line), obtained from $C_{slow,2}(t)$, is narrower than $P_{slow,1}(k)$ and centered approximately at $\langle k \rangle$. Interestingly, when both *CV1* and *CV2* are treated as slow variables, $P_{slow,1\text{-}2}(k)$ (green line) shows reduced weight on the large-$k$ side and increased weight on the small-$k$ side (e.g., $k \approx 2\times10^{-3}$ and $k \approx 4\times10^{-4}$, respectively, for particle A at $T = 0.511$) of the main peak of $P(k)$. This redistribution of weight indicates that incorporating both *CV1* and *CV2* as slow variables provides a more appropriate approximation of $C_S(t)$ over the entire timescale than assuming *CV1* alone as the slow variable. Finally, at the lower temperature $T = 0.455$, these two variables are insufficient. However, adding *CV2* to *CV1* induces a similar $k$-dependent redistribution of weight as in the case of $T = 0.511$, thereby improving the approximation of the survival probability.

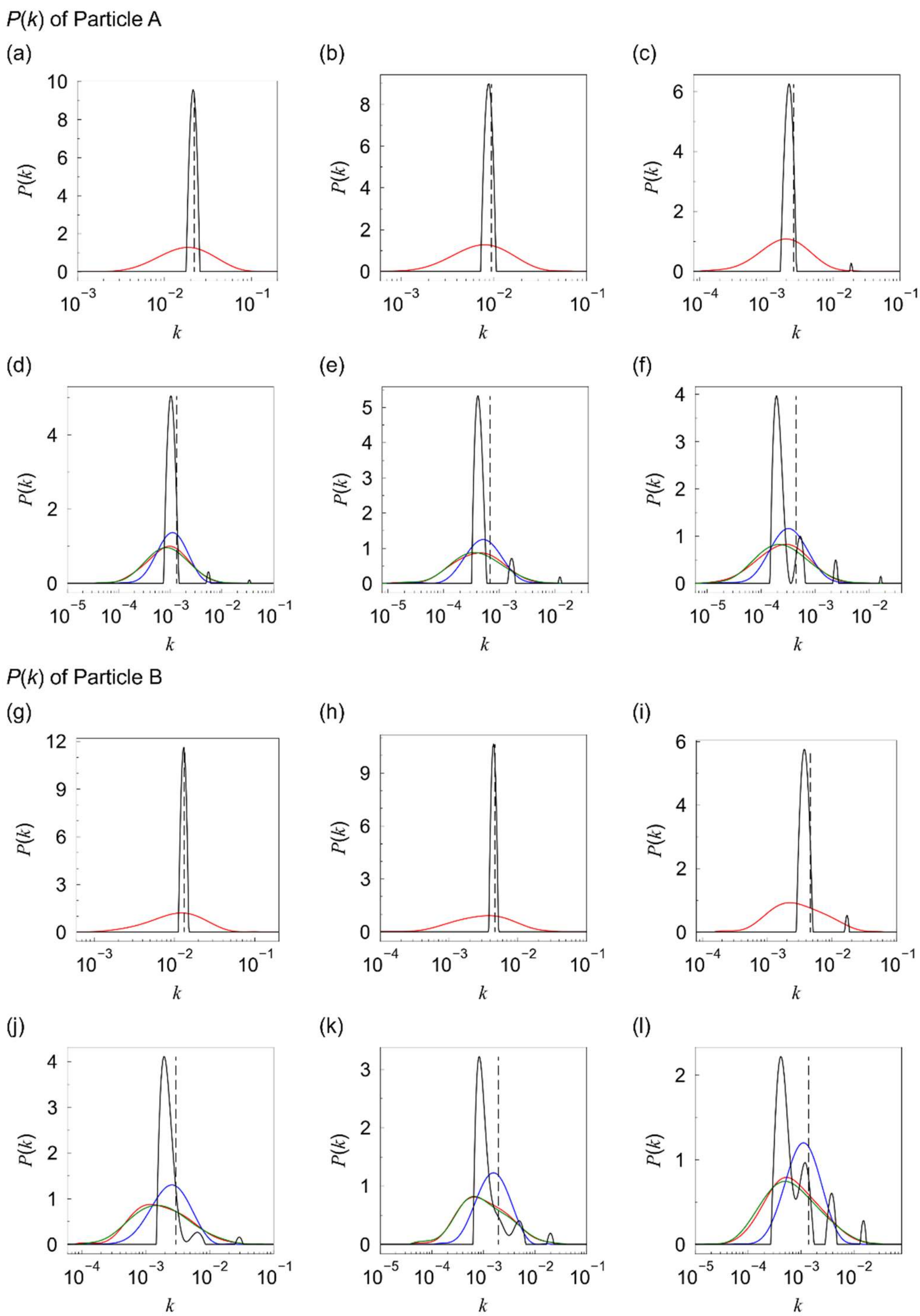

Figure 9. Rate distributions $P(k)$ (black), $P_{\mathrm{slow},1}(k)$ (red), $P_{\mathrm{slow},2}(k)$ (blue), and $P_{\mathrm{slow},1\text{-}2}(k)$ (green) for particle A (a-f) and particle B (g-l) obtained from the inverse Laplace transform of $C_S(t)$, $C_{\mathrm{slow},1}(t)$,

$C_{\mathrm{slow},2}(t)$, and $C_{\mathrm{slow},1\text{-}2}(t)$, respectively, at (a) and (g) $T = 0.982$, (b) and (h) $T = 0.698$, (c) and (i) $T = 0.550$, (d) and (j) $T = 0.511$, (e) and (k) $T = 0.476$, and (f) and (l) $T = 0.455$. The vertical dashed lines indicate the values of $\langle k \rangle$.

At the end of this subsection, we examined the difference in the influence of *CV1* and *CV2* on the jump dynamics of particles A and B. Figure 10 shows that the $D_{\mathrm{eq}}$ values of *CV1*, *CV2*, and *CV1-2* for particle A are 0.595, 0.309, and 0.619, respectively, whereas the corresponding values for particle B are 0.698, 0.264, and 0.767, respectively. Thus, when both *CV1* and *CV2* are treated as slow variables, $D_{\mathrm{eq}}$ increases by 4.1% for particle A compared with *CV1* alone, whereas it increases by 10.0% for particle B. These results indicate that particles in the first and second coordination shells around the jumping particle B undergo stronger cooperative structural changes than those around the jumping particle A (Fig. S16). Notably, this behavior regarding the cooperativity between *CV1* and *CV2* below $T = 0.550$ is distinct from the result that the jump motions of particle A are more spatially extended than those of particle B (Fig. 2). Furthermore, the difference in cooperativity between the two species is also reflected in the more pronounced increase at small $k$ and decrease at large $k$ in $P_{\mathrm{slow},1\text{-}2}(k)$ (green) relative to $P_{\mathrm{slow},1}(k)$ (red) for particle B than for particle A [Figs. 9 (d), (e), and (f) compared with Figs. 9 (j), (k), and (l)].

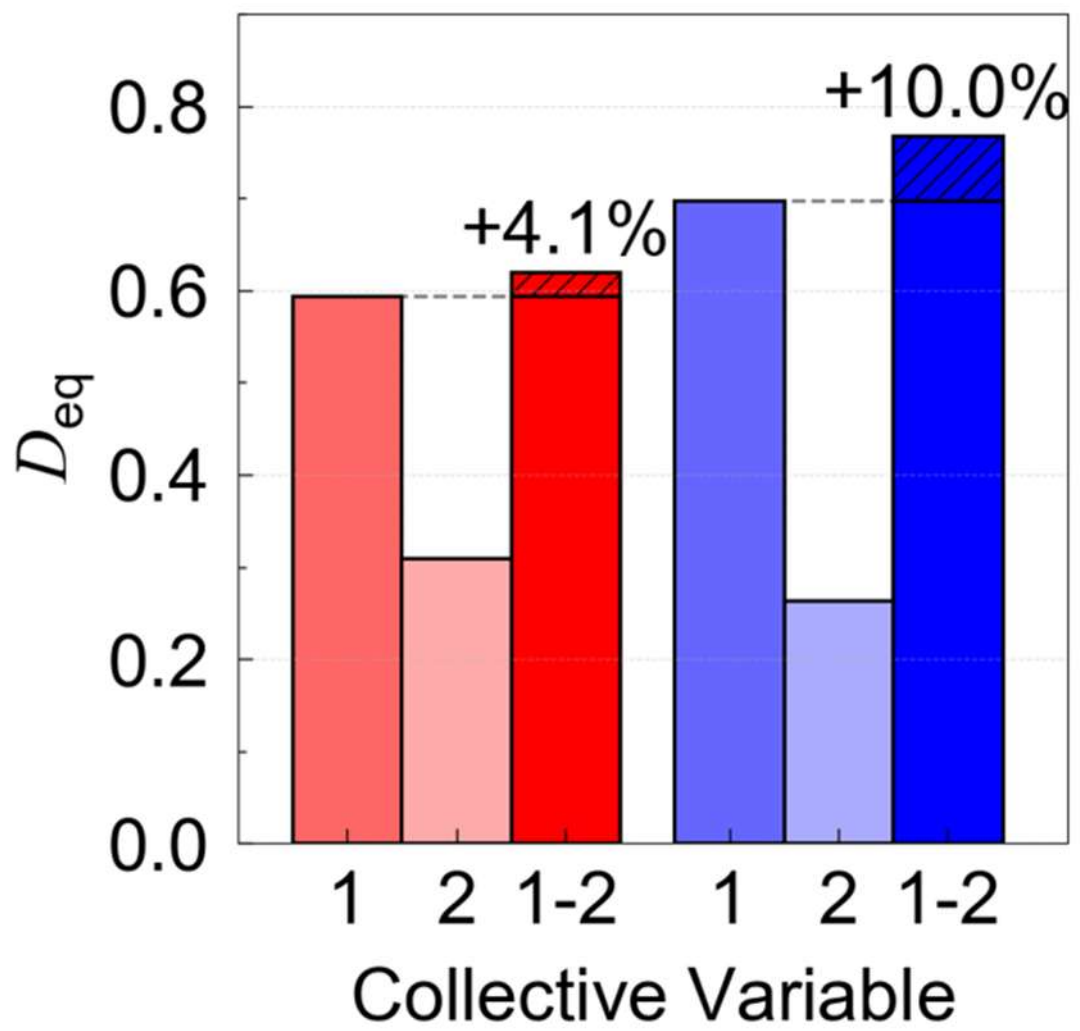

Figure 10. Comparison of $D_{\mathrm{eq}}$ for the collective variables, *CV1*, *CV2*, and *CV1-2*, for particles A (red) and B (blue) at $T = 0.455$.

### D. Connection to the static correlation length

We next examined the correspondence between the slow variables identified above and the PTS static correlation length, $\xi_{\mathrm{pts}}$. This quantity has been extensively studied in the KALJ model[105-107, 120]

First, following previous studies, we calculated $\xi_{pts}$ without distinguishing between central particles A and B. As a result, $\xi_{pts}$ increases from 0.7 to 2.4 as the temperature decreases from $T = 0.982$ to 0.455, in quantitative agreement with previous work [Figs. 11 (a), and S17 (a)–S17 (c)].[105, 106] Furthermore, we confirmed the result of the previous work[105] that the increase in $\xi_{pts}$ upon cooling correlates with the dynamic slowdown characterized by $\tau_\alpha$ [Fig. S17 (d)]. Thus, the continuous and substantial increase in $\xi_{pts}$ across the entire temperature range is characteristic of fragile liquids that exhibit super-Arrhenius temperature dependence of their dynamic slowdown. In contrast, in strong liquids, including silica melt, $\xi_{pts}$ grows more gradually and eventually saturates.[35, 101]

Next, we analyzed the static correlation lengths, $\xi_{pts,A}$ and $\xi_{pts,B}$, by distinguishing whether the central particle is A or B [Fig. 11 (a)]. We first discuss $\xi_{pts,A}$. At $T = 0.982$, $\xi_{pts,A}$ is shorter than the onset distance of the RDF, suggesting that particle A moves independently. This result is consistent with the finding that no slow variables are involved in $C_S(t)$ at this temperature. As the temperature decreases to 0.698, $\xi_{pts,A}$ increases to ~1.2, entering the region corresponding to *CV1*. This is consistent with the earlier result showing that this variable contributes to $C_S(t)$. As the temperature further decreases to 0.511, $\xi_{pts,A}$ extends into the region corresponding to *CV2*, indicating that this range is involved in the jump dynamics. At $T = 0.455$, $\xi_{pts,A}$ extends beyond the range of *CV2*, suggesting that even more distant regions contribute to the jump dynamics. Qualitatively similar behavior is observed for $\xi_{pts,B}$; that is, $\xi_{pts,B}$ also grows into the regions corresponding to *CV1* and *CV2* at the same temperatures as $\xi_{pts,A}$ (Fig. 11). These results suggest an apparent correspondence between the static correlation length and the spatial extent of the important slow structural variables revealed by the survival probabilities. A similar relationship between these quantities has also been found in silica melt.[101] However, the correlation between the spatial extent of the slow variables revealed by the survival probability and the growth of the PTS correlation length does not necessarily hold universally. For example, in a polydisperse hard-disk system with hexatic order, an earlier study reported that the PTS correlation length grows much more weakly than the bond-orientational correlation length.[121] Meanwhile, a later study attributed this discrepancy to a mismatch in packing fraction and polydispersity between the bulk and the cavity used in the PTS calculation and supported the robustness of the PTS method in these systems with medium-range crystalline order.[122]

Finally, we examined the effects of the collective variables *CV1* and *CV2*, identified earlier as slow variables, on the local structural relaxation of jumping particles by performing distance-constrained MC simulations and comparing them with reference fast variables at $T = 0.511$. In this analysis, we calculated the overlap function defined only for jumping particles, using configurations extracted from MD trajectories before jump events (see Supporting Material Texts for details). As shown in Figs. 11 (b) and 11 (c), the overlap functions under the constraints of *CV1* and *CV2'* exhibit the strongest and second-strongest slowdown effects among the candidate one-dimensional variables, consistent with the results presented earlier. Here, *CV2'* denotes a three-particle subset of *CV2* consisting of the 31st to 33rd (25th to 27th) nearest neighbors for a jumping particle A (B). This indicates that constraining *CV1* and *CV2'* hinders the structural preparation preceding a jump, thereby slowing the motion of the jumping particle. In contrast, constraining the particle set in the inner region of the first coordination shell (*Ref1*) produces a much weaker slowdown. Here, *Ref1* consists of the 2nd to 6th (2nd to 4th) nearest neighbors for particle A (B). Furthermore, we examined the effects of constraining the particle set in the inner region of the second coordination shell (*Ref2'*), where *Ref2'* consists of the 20th to 22nd (14th to 16th) nearest neighbors for particle A (B). We found that the slowdown remains merely comparable to that observed for the more distant *CV2'*, indicating that *Ref2'* is less involved in the jump dynamics than *CV2'*. Because these reference variables relax very rapidly, their contribution to the structural preparation for a jump is small; consequently, their effect on the jumping particle is limited compared with that of constraining *CV1* or *CV2'*. The overlap function under the constraint of the two-dimensional slow variable *CV1-2'* exhibits a more pronounced slowdown than that under the constraints of *CV1* alone or *Ref1-2'*, constructed from two non-slow variables. These results indicate that structural relaxation of the jumping particle is significantly suppressed when the identified slow variables are constrained.

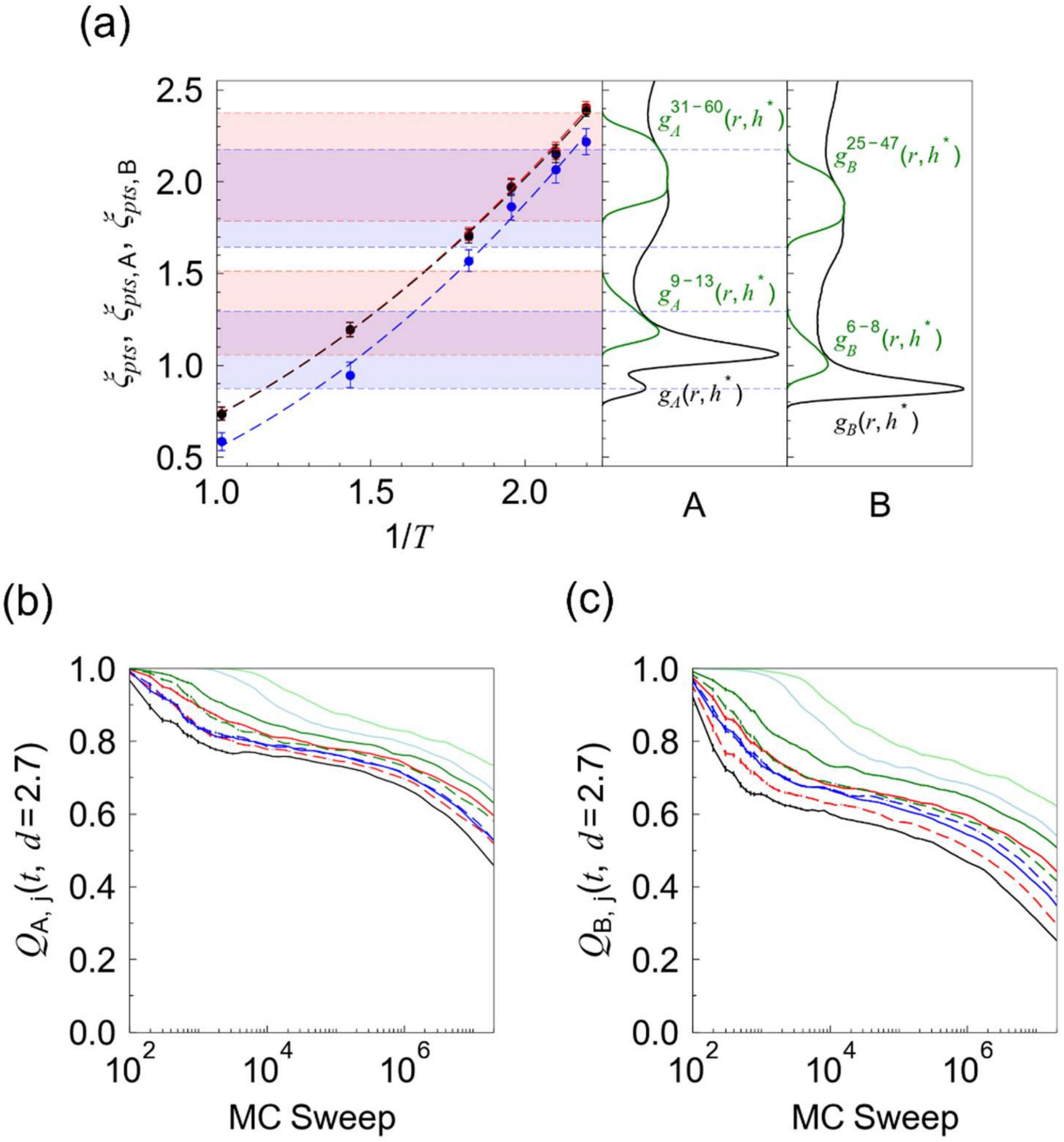

Figure 11. (a) Point-to-set static correlation length, $\xi_{\mathrm{pts}}$, for particles A (red), B (blue), and both species (black). (b) Block-averaged overlap functions of particle A, $Q_{A,\mathrm{j}}(t,d=2.7)$, along MC sweeps at $T$ = 0.511 calculated using the MC simulations with constraints on the distances to neighboring particles. (c) The same as (b) for particle B. In (a), dashed curves are shown as guides to the eye. Error bars represent standard deviations evaluated using the bootstrap method. Green curves in the right panels display the radial distributions restricted to the labeled neighboring particles, $g_{\mathrm{A}}^{9\text{-}13}(r,h^{*})$, $g_{\mathrm{A}}^{31\text{-}60}(r,h^{*})$, $g_{\mathrm{B}}^{6\text{-}8}(r,h^{*})$, and $g_{\mathrm{B}}^{25\text{-}47}(r,h^{*})$, at $T$ = 0.455. The black curves in the right panels display $g_{\mathrm{A}}(r,h^{*})$ and $g_{\mathrm{B}}(r,h^{*})$, respectively. In (b) and (c), colors represent the unconstrained case (black), and cases with additional distance constraints applied to *CV1* (solid red), *CV2* (solid light blue), *CV2'* (solid blue), *CV1-2* (solid light green), *CV1-2'* (solid green), *Ref1* (dashed red), *Ref2'* (dashed blue), and *Ref1-2'* (dashed green). The data are block-averaged using a bin width of 0.6 on the logarithmic scale, and error bars represent the corresponding standard deviations.

## IV. Conclusions

We investigated the microscopic mechanisms underlying the super-Arrhenius dynamic slowdown in the fragile KALJ liquid by analyzing particle jump motions. Using the hop function, which measures the displacement of jumping particles, as the reaction coordinate, we demonstrated the progressive emergence of the non-Poissonian dynamics that give rise to dynamic disorder in the supercooled KALJ liquid.

At the temperature immediately below the onset of supercooling, we found that neighboring particles in the outer regions of the first coordination shell exhibit significant shifts in their distance distributions during jump motions. By examining the survival probability of the cage state, we identified the distance distributions of these neighbors as the slow variables that induce the fluctuations in the jump rate $k(t)$. As the temperature further decreases, the relevant slow variables extend to the outer regions of the second coordination shell and beyond. These slow variables modulate the jump dynamics and expand spatially in parallel with the growth of the static correlation length, providing a consistent picture of the characteristic super-Arrhenius dynamic slowdown in the fragile supercooled liquids. Furthermore, we revealed that the jump motion of particle B involves more strongly cooperative displacements between the first and second coordination shells than that of particle A, highlighting species-dependent differences in the underlying local structural rearrangements associated with jump motions.

Compared with tetrahedral network-forming liquids, such as water and silica melt, we found that defining effective slow variables in the KALJ model is more challenging because of its higher coordination number. To address this difficulty, we identified relevant variables using the average distances of selected groups of nearest neighboring particles associated with large $D_{\mathrm{eq}}$ values. In future work, we plan to pursue a machine learning-based approach to extract collective slow variables more efficiently.

## SUPPLEMENTARY MATERIAL

See the supplementary material for additional details referenced in the text.

**Acknowledgments**

The computation was performed using Research Center for Computational Science, Okazaki, Japan (Projects: 24-IMS-C193 and 25-IMS-C223). This work was supported in part by JSPS KAKENHI (Grant-in-Aid for Scientific Research A) (JP21H04676), Challenging Research (Pioneering) (JP23K17361) and a research grant from DAICEL.

**AUTHOR DECLARATIONS**

Conflict of Interest

The authors have no conflicts to disclose.

**Author Contributions**

**Zhiye Tang**: Conceptualization (equal); Data curation (equal); Formal analysis (equal); Investigation (equal); Methodology (equal); Software (equal); Validation (equal); Visualization (equal); Writing – original draft (equal); Writing – review & editing (equal). **Shubham Kumar**: Conceptualization (equal); Formal analysis (equal); Methodology (equal); Writing – review & editing (equal); Investigation (equal); Validation (equal). **Shinji Saito**: Conceptualization (equal); Formal analysis (equal); Funding acquisition (equal); Investigation (equal); Methodology (equal); Project administration (equal); Resources (equal); Supervision (lead); Validation (equal); Writing – original draft (equal); Writing – review & editing (equal).

**DATA AVAILABILITY**

The data that support the findings of this study are available within the article and its supplementary material.

## References

1. G. Biroli and J. P. Garrahan, J. Chem. Phys. **138**, 12a301 (2013).
2. P. G. Debenedetti and F. H. Stillinger, Nature **410,** 259-267 (2001).
3. M. Ediger, Annu. Rev. Phys. Chem. **51,** 99-128 (2000).
4. H. Tanaka, J. Phys. Chem. B **129,** 789-813 (2025).
5. M. Ciamarra, R. Pastore and A. Coniglio, Soft Matter **12,** 358-366 (2016).
6. V. Dubey, S. Dueby and S. Daschakraborty, Phys. Chem. Chem. Phys. **23,** 19964-19986 (2021).
7. C. Angell, J. Non-Cryst. Solids **131,** 13-31 (1991).
8. C. Angell, Chem. Rev. **102,** 2627-2649 (2002).
9. J. Horbach and W. Kob, Phys. Rev. B **60,** 3169-3181 (1999).
10. P. G. Debenedetti, T. M. Truskett and C. P. Lewis, Adv. Chem. Eng. **28,** 21-79 (2001).
11. G. Ruocco, F. Sciortino, F. Zamponi, C. De Michele and T. Scopigno, J. Chem. Phys. **120,** 10666-10680 (2004).
12. G. Sun, L. Li and P. Harrowell, J. Non-Cryst. Solids X **13**, 100080 (2022).
13. M. D. Ediger, C. A. Angell and S. R. Nagel, J. Phys. Chem. **100,** 13200-13212 (1996).
14. J. Dyre, Rev. Mod. Phys. **78,** 953-972 (2006).
15. M. Ediger and P. Harrowell, J. Chem. Phys. **137**, 080901 (2012).
16. G. Fredrickson and H. Andersen, Phys. Rev. Lett. **53,** 1244-1247 (1984).
17. Y. Jung, J. Garrahan and D. Chandler, J. Chem. Phys. **123**, 084509 (2005).
18. L. Berthier and G. Biroli, Rev. Mod. Phys. **83,** 587-645 (2011).
19. T. Kirkpatrick, Phys. Rev. A **31,** 939-944 (1985).
20. W. Kob, Supercooled Liquids **676,** 28-44 (1997).
21. L. M. C. Janssen, Front. Phys. **6**, 97 (2018).
22. G. Adam and J. H. Gibbs, The Journal of Chemical Physics **43,** 139-146 (1965).
23. T. R. Kirkpatrick, D. Thirumalai and P. G. Wolynes, Phys. Rev. A **40,** 1045-1054 (1989).
24. X. Xia and P. G. Wolynes, Proc. Natl. Acad. Sci. U.S.A. **97,** 2990-2994 (2000).
25. C. Royall, S. Williams, T. Ohtsuka and H. Tanaka, Nat. Mat. **7,** 556-561 (2008).
26. A. Malins, J. Eggers, H. Tanaka and C. Royall, Faraday Discuss. **167,** 405-423 (2013).
27. H. Tong and H. Tanaka, Nat. Comm. **10**, 5596 (2019).
28. J. Bouchaud and G. Biroli, J. Chem. Phys. **121,** 7347-7354 (2004).
29. A. Montanari and G. Semerjian, J. Stat. Phys. **125,** 22-54 (2006).
30. A. Cavagna, T. Grigera and P. Verrocchio, Phys. Rev. Lett. **98**, 187801 (2007).
31. L. Berthier, G. Biroli, J. P. Bouchaud and R. L. Jack, *Overview of different characterisations of dynamic heterogeneity*. (Oxford University Press, 2011).
32. L. Berthier and D. Reichman, Nat. Rev. Phys. **5,** 102-116 (2023).
33. L. Berthier, P. Charbonneau, D. Coslovich, A. Ninarello, M. Ozawa and S. Yaida, Proc. Natl. Acad. Sci. U.S.A. **114,** 11356-11361 (2017).
34. S. Karmakar, C. Dasgupta and S. Sastry, Rep. Prog. Phys. **79**, 016601 (2016).
35. I. Tah and S. Karmakar, Phys. Rev. Mat. **6**, 035601 (2022).
36. P. Steinhardt, D. Nelson and M. Ronchetti, Phys. Rev. B **28,** 784-805 (1983).
37. G. Tarjus, S. Kivelson, Z. Nussinov and P. Viot, J. Phys.: Condens. Matter **17,** R1143-R1182 (2005).
38. P. Crowther, F. Turci and C. Royall, J. Chem. Phys. **143**, 044503 (2015).
39. C. Royall and S. Williams, Phys. Rep. **560,** 1-75 (2015).
40. D. Coslovich and R. Jack, J. Stat. Mech.: Theory Exp. 074012 (2016).
41. H. Tanaka, H. Tong, R. Shi and J. Russo, Nat. Rev. Phys. **1,** 333-348 (2019).
42. M. Rahman, B. Carter, S. Saw, I. Douglass, L. Costigliola, T. Ingebrigtsen, T. Schroder, U. Pedersen and J. Dyre, Molecules **26**, 1746 (2021).
43. F. Sciortino, W. Kob and P. Tartaglia, J. Phys.: Condens. Matter **12,** 6525-6534 (2000).

44. E. La Nave, S. Sastry and F. Sciortino, Phys. Rev. E **74**, 050501 (2006).
45. Z. Raza, B. Alling and I. A. Abrikosov, J. Phys.: Condens. Matter **27**, 293201 (2015).
46. M. Sampoli, P. Benassi, R. Eramo, L. Angelani and G. Ruocco, J. Phys.: Condens. Matter **15,** S1227-S1236 (2003).
47. V. Novikov, Y. Ding and A. Sokolov, Phys. Rev. E **71**, 061501 (2005).
48. A. Widmer-Cooper, H. Perry, P. Harrowell and D. R. Reichman, Nat. Phys. **4,** 711-715 (2008).
49. D. Coslovich and G. Pastore, J. Phys.: Condens. Matter **21**, 285107 (2009).
50. L. Yan, G. Düring and M. Wyart, Proc. Natl. Acad. Sci. U.S.A. **110,** 6307-6312 (2013).
51. E. Weeks, J. Crocker, A. Levitt, A. Schofield and D. Weitz, Science **287,** 627-631 (2000).
52. P. Chaudhuri, L. Berthier and W. Kob, Phys. Rev. Lett. **99**, 060604 (2007).
53. R. Candelier, A. Widmer-Cooper, J. Kummerfeld, O. Dauchot, G. Biroli, P. Harrowell and D. Reichman, Phys. Rev. Lett. **105**, 135702 (2010).
54. S. Dueby, V. Dubey and S. Daschakraborty, J. Phys. Chem. B **123,** 7178-7189 (2019).
55. T. Kikutsuji, K. Kim and N. Matubayasi, J. Chem. Phys. **150**, 204502 (2019).
56. R. Pastore, T. Kikutsuji, F. Rusciano, N. Matubayasi, K. Kim and F. Greco, J. Chem. Phys. **155**, 114503 (2021).
57. O. Rubner and A. Heuer, Phys. Rev. E **78**, 011504 (2008).
58. J. Helfferich, F. Ziebert, S. Frey, H. Meyer, J. Farago, A. Blumen and J. Baschnagel, Phys. Rev. E **89**, 042603 (2014).
59. R. Pastore, A. Coniglio and M. Ciamarra, Sci. Rep. **5**, 11770 (2015).
60. S. Saito, J. Chem. Phys. **160**, 194506 (2024).
61. R. Zwanzig, Acc. Chem. Res. **23,** 148-152 (1990).
62. J. Wang and P. Wolynes, Phys. Rev. Lett. **74,** 4317-4320 (1995).
63. V. Chernyak, M. Schulz and S. Mukamel, J. Chem. Phys. **111,** 7416-7425 (1999).
64. J. Cao, Chem. Phys. Lett. **327,** 38-44 (2000).
65. Y. Gebremichael, M. Vogel and S. Glotzer, J. Chem. Phys. **120,** 4415-4427 (2004).
66. V. Lubchenko and P. G. Wolynes, Annu. Rev. Phys. Chem. **58,** 235-266 (2007).
67. M. Nandi and S. Bhattacharyya, Phys. Rev. Lett. **126**, 208001 (2021).
68. M. Sharma, M. Nandi and S. Bhattacharyya, Phys. Rev. E **105**, 044604 (2022).
69. S. Schoenholz, E. Cubuk, D. Sussman, E. Kaxiras and A. Liu, Nat. Phys. **12,** 469 (2016).
70. X. Ma, Z. Davidson, T. Still, R. Ivancic, S. Schoenholz, A. Liu and A. Yodh, Phys. Rev. Lett. **122**, 028001 (2019).
71. E. Cubuk, S. Schoenholz, E. Kaxiras and A. Liu, J. Phys. Chem. B **120,** 6139-6146 (2016).
72. V. Bapst, T. Keck, A. Grabska-Barwinska, C. Donner, E. Cubuk, S. Schoenholz, A. Obika, A. Nelson, T. Back, D. Hassabis and P. Kohli, Nat. Phys. **16,** 448 (2020).
73. E. Boattini, F. Smallenburg and L. Filion, Phys. Rev. Lett. **127**, 088007 (2021).
74. H. Shiba, M. Hanai, T. Suzumura, T. Shimokawabe and T. Shimokawabe, J. Chem. Phys. **158**, 084503 (2023).
75. G. Jung, G. Biroli and L. Berthier, Phys. Rev. Lett. **130**, 238202 (2023).
76. G. Jung, R. Alkemade, V. Bapst, D. Coslovich, L. Filion, F. Landes, A. Liu, F. Pezzicoli, H. Shiba, G. Volpe, F. Zamponi, L. Berthier and G. Biroli, Nat. Rev. Phys. **7,** 91-104 (2025).
77. W. Kob and H. C. Andersen, Phys. Rev. Lett. **73,** 1376-1379 (1994).
78. W. Kob and H. Andersen, Phys. Rev. E **51,** 4626-4641 (1995).
79. W. Kob, C. Donati, S. Plimpton, P. Poole and S. Glotzer, Phys. Rev. Lett. **79,** 2827-2830 (1997).
80. W. Kob and J. Barrat, Phys. Rev. Lett. **78,** 4581-4584 (1997).
81. C. Donati, S. Glotzer, P. Poole, W. Kob and S. Plimpton, Phys. Rev. E **60,** 3107-3119 (1999).
82. T. Middleton, J. Hernández-Rojas, P. Mortenson and D. Wales, Phys. Rev. B **64**, 184201 (2001).
83. S. Ashwin and S. Sastry, J. Phys.: Condens. Matter **15,** S1253-S1258 (2003).
84. L. Berthier and W. Kob, J. Phys.: Condens. Matter **19**, 205130 (2007).

85. R. Stein and H. Andersen, Phys. Rev. Lett. **101**, 267802 (2008).
86. R. Brüning, D. St-Onge, S. Patterson and W. Kob, J. Phys.: Condens. Matter **21**, 035117 (2009).
87. U. Nandi, A. Banerjee, S. Chakrabarty and S. Bhattacharyya, J. Chem. Phys. **145**, 034503 (2016).
88. U. Pedersen, T. Schroder and J. Dyre, Phys. Rev. Lett. **120**, 165501 (2018).
89. Q. Yuan, X. Xu, J. Douglas and W. Xu, J. Phys. Chem. B **128,** 9889-9904 (2024).
90. Y. Yang, Y. Lu and L. An, J. Phys. Chem. B **129,** 12774–12781 (2025).
91. S. Kumawat, M. Sharma, U. K. Nandi, I. Tah and S. M. Bhattacharyya, J. Chem. Phys. **163**, 204505 (2025).
92. A. Thompson, H. Aktulga, R. Berger, D. Bolintineanu, W. Brown, P. Crozier, P. Veld, A. Kohlmeyer, S. Moore, T. Nguyen, R. Shan, M. Stevens, J. Tranchida, C. Trott and S. Plimpton, Comput. Phys. Commun. **271**, 108171 (2022).
93. R. Pastore, A. Coniglio and M. Ciamarra, Soft Matter **10,** 5724-5728 (2014).
94. R. Candelier, O. Dauchot and G. Biroli, Phys. Rev. Lett. **102**, 088001 (2009).
95. A. Smessaert and J. Rottler, Phys. Rev. E **88**, 022314 (2013).
96. M. Schnitzer and S. Block, Cold Spring Harb. Symp. Quant. Biol. **60,** 793-802 (1995).
97. J. Moffitt and C. Bustamante, FEBS J. **281,** 498-517 (2014).
98. B. Halle and F. Persson, J. Chem. Theory Comput. **9,** 2838-2848 (2013).
99. Y. Matsumura and S. Saito, J. Chem. Phys. **154**, 224113 (2021).
100. D. Chandler and J. Garrahan, Annu. Rev. Phys. Chem. **61,** 191-217 (2010).
101. S. Kumar, Z. Tang and S. Saito, J. Chem. Phys. **164**, 024503 (2026).
102. S. Kullback and R. A. Leibler, Ann. Math. Stat. **22,** 79-86 (1951).
103. G. Biroli, J. Bouchaud, A. Cavagna, T. Grigera and P. Verrocchio, Nat. Phys. **4,** 771-775 (2008).
104. L. Berthier and W. Kob, Phys. Rev. E **85**, 011102 (2012).
105. G. Hocky, T. Markland and D. Reichman, Phys. Rev. Lett. **108**, 225506 (2012).
106. Y. Li, W. Xu and Z. Sun, J. Chem. Phys. **140**, 124502 (2014).
107. L. Berthier, P. Charbonneau and S. Yaida, J. Chem. Phys. **144**, 024501 (2016).
108. B. Charbonneau, P. Charbonneau and G. Tarjus, Phys. Rev. Lett. **108**, 035701 (2012).
109. G. Biroli and C. Cammarota, Phys. Rev. X **7**, 011011 (2017).
110. V. Levashov, Phys. Rev. E **111**, 035403 (2025).
111. H. Tong and H. Tanaka, Phys. Rev. Lett. **124**, 225501 (2020).
112. A. Keys, L. Hedges, J. Garrahan, S. Glotzer and D. Chandler, Phys. Rev. X **1**, 021013 (2011).
113. P. Das and S. Sastry, J. Non-Cryst. Solids X **14,** 100098 (2022).
114. L. Berthier and G. Tarjus, Phys. Rev. Lett. **103**, 170601 (2009).
115. K. Kim and S. Saito, J. Chem. Phys. **138**, 12A506 (2013).
116. F. Stillinger, Phys. Rev. B **41,** 2409-2416 (1990).
117. D. Johnston, Phys. Rev. B **74**, 184430 (2006).
118. A. Krishnan and B. Epureanu, Bull. Math. Biol. **73,** 2452-2482 (2011).
119. Y. Song, L. Venkataramanan, M. Hürlimann, M. Flaum, P. Frulla and C. Straley, J. Magn. Reson. **154,** 261-268 (2002).
120. B. Mei, Z. Wang, Y. Lu, H. Li and L. An, J. Chem. Phys. **147**, 114507 (2017).
121. J. Russo and H. Tanaka, Proc. Natl. Acad. Sci. U.S.A. **112,** 6920-6924 (2015).
122. I. Tah, S. Sengupta, S. Sastry, C. Dasgupta and S. Karmakar, Phys. Rev. Lett. **121**, 085703 (2018).

# Supporting Material for Super-Arrhenius Dynamic Slowdown Revealed by Slow Variable Modulation in the Fragile Supercooled Liquid

Zhiye Tang[1,2], Shubham Kumar[1], and Shinji Saito[1,2,*]

[1]Institute for Molecular Science, Myodaiji, Okazaki, Aichi 444-8585, Japan

[2]The Graduate University for Advanced Studies (SOKENDAI), Myodaiji, Okazaki, Aichi 444-8585, Japan

*Corresponding author: shinji@ims.ac.jp

Supporting Material Texts

**Monte Carlo (MC) Simulation with Constraints on Distance to Particles**

We investigated the extent to which jump motion depends on the collective rearrangement of the surrounding particles by performing MC simulations with constraints imposed on the distances between the jumping particle and its surrounding particles. The degree to which the motion of the jumping particle is suppressed under constrained conditions was quantitatively evaluated using an overlap function.

**Definition of surrounding particles**

To identify which regions of the local environment are important for the jump motion of the central particle, we classified the surrounding particles into four sets and compared the motion of the jumping particle when each set was constrained. As already shown in the main text (Fig. 6), the particles located in the outer regions of the first and second coordination shells of the jumping particle (denoted as *CV1* and *CV2*, respectively) exhibit large $D_{eq}$ values. In contrast, the particles located closer to the inner parts of these coordination shells (hereafter referred to as *Ref1* and *Ref2*) show relatively small $D_{eq}$ values. We therefore constrained the particles belonging to *CV1*, *CV2*, *Ref1*, and *Ref2*, respectively, and examined how the displacement of the central jumping particle (i.e., the degree of suppression of its motion) changes.

The number of particles included in *CV1* is five when the jumping particle is particle *A*, and three when it is particle *B*. To ensure a fair comparison with *CV1*, we selected the same number of particles for *Ref1* for particles A and B. The particle lists for *CV1* and *Ref1* are provided in Table S1. In contrast, the number of particles belonging to *CV2* is much larger: 30 when the jumping particle is *A* and 23 when it is *B*, respectively. However, because the number of particles located in the region of *Ref2* is smaller than these values, we defined a reduced set, *CV2'*, consisting of only the three particles in *CV2* that are closest to the jumping particle. Similarly, for the reference set located closer to the inner part of the second coordination shell, we defined *Ref2'* as the set of the three particles closest to the jumping particle. The particle lists for *CV2*, *CV2'*, and *Ref2'* for both particle *A* and particle *B* are given in Table S1.

**Distance-constrained MC simulations**

For the initial configurations of the distance-constrained MC simulations, we used structures extracted from molecular dynamics (MD) trajectories immediately before jump events. Specifically, for both particle A and particle B, the configuration at $10\Delta t$ before the jump time was employed as the initial configuration, where $\Delta t$ denotes the time scale used in the definition of the hopping function. To ensure statistical independence, for each initial configuration, 1280 configurations for particles A and B were selected from MD trajectories and were separated by at least $2\tau_{\alpha,A}$ and $2\tau_{\alpha,B}$, respectively. Here, $\tau_{\alpha,A}$ and $\tau_{\alpha,B}$ denote the α-relaxation times for particles A and B, respectively.

For each initial configuration, the system was arranged such that the particle just before the jump was located at the center of a cubic cavity with side length $d = 2.7$, and all particles outside the cavity were pinned. The cavity size was chosen to suppress extended bulk-like rearrangements while retaining sufficient space for local relaxation of the jumping particle and its surrounding neighbors.

In each MC sweep, standard trial moves for all unpinned particles were performed, and the constraints were subsequently applied as follows. In the first phase, each unpinned particle was displaced in a random direction by a distance smaller than a prescribed maximum value (0.3 in reduced length units), and the trial configuration was subjected to the Metropolis acceptance test. The order of particle updates was also randomized. After completion of all standard MC moves in a sweep, the nearest-neighbor list was rebuilt from the updated trial configuration. In the second phase, the constrained neighbors of the jumping particle were displaced by rescaling their distances from the jumping particle to the prescribed initial values, thereby enforcing the constraints. This procedure was carried out sequentially for each constrained neighboring particle, with each constraint-enforcing trial move again subjected to a Metropolis acceptance test. If any constraint move was rejected or altered the ordering of the nearest-neighbor list, the configuration was reverted to that at the beginning of the current sweep; otherwise, the updated configuration was retained for the next sweep.

**Quantifying jump motion using the overlap function**

To examine how the motion of a jump particle is altered under constraints imposed on the surrounding particles, we analyzed the time evolution of an overlap function defined only for the jump particle. This overlap function was defined as being 1 when the jump particle remained within its initial grid cell and 0 when it moved out of that cell. Therefore, a slower decay of the overlap function indicates that the displacement of the jump particle is more strongly suppressed by the distance constraint.

In this MC scheme, the neighboring particles subject to the constraint remain mobile, and the particle indices defining them are allowed to change freely. However, because changes in the distances between these particles and the jumping particle are restricted relative to the unconstrained case, collective rearrangements accompanied by distance changes are suppressed. Therefore, if such rearrangements are required for the structural relaxation preceding a jump, the distance constraint imposed here is expected to delay the jump dynamics. In this study, we analyzed this scheme at $T = 0.511$.

Table S1. Nearest neighboring particle lists.

| | Particle A | Particle B |
|---|---|---|
| *CV1* | 9th–13th | 6th–8th |
| *CV2* | 31st–60th | 25th–47th |
| *CV2'* | 31st–33rd | 25th–27th |
| *Ref1* | 2nd–6th | 2nd–4th |
| *Ref2'* | 20th–22nd | 14th–16th |

Table S2. $\Delta t$ values used in the hop function for particles A and B.

| $T$ | 0.982 | 0.698 | 0.550 | 0.511 | 0.476 | 0.455 |
|---|---|---|---|---|---|---|
| $\Delta t$ (A) | 0.58 | 1.08 | 1.67 | 1.72 | 1.72 | 1.72 |
| $\Delta t$ (B) | 0.61 | 0.79 | 1.61 | 1.68 | 1.71 | 1.72 |

Table S3. Replica parameters for $T = 0.982$.

| $d$ | 1.3 | 1.5 | 1.7 | 1.9 | 2.1 | 2.3 | 2.5 | 2.7 | 2.9 |
|---|---|---|---|---|---|---|---|---|---|
| $S_{\mathrm{EQ}}$ | 1000 | 1000 | 1000 | 1000 | 1000 | 1000 | 1000 | 3000 | 3000 |
| $S_{\mathrm{PROD}}$ | 4000 | 4000 | 4000 | 4000 | 4000 | 7000 | 9000 | 7000 | 7000 |
| $T_{\mathrm{DEC}}$ | 1.0 | 1.0 | 1.0 | 1.0 | 1.0 | 1.0 | 1.0 | 1.0 | 1.0 |
| $\lambda_{\mathrm{DEC}}$ | 0.80 | 0.80 | 0.80 | 0.80 | 0.80 | 0.80 | 0.80 | 0.80 | 0.80 |
| $N_{\mathrm{REP}}$ | 5 | 5 | 7 | 9 | 10 | 11 | 12 | 13 | 16 |

* $S_{\mathrm{EQ}}$ and $S_{\mathrm{PROD}}$ denote the number of sweeps for equilibration and production, respectively, between two consecutive attempts of Hamiltonian exchanges. $T_{\mathrm{DEC}}$ and $\lambda_{\mathrm{DEC}}$ indicate the decoupling temperature and scaling factor of $\sigma_{\alpha\beta}$ in the Lennard-Jones potential for a pair of particles $\alpha$ and $\beta$ [Eq. (1)], respectively. For a Monte Carlo calculation with $N_{\mathrm{REP}}$ replicas, the temperature $T_a$ and scaling factor $\lambda_a$ of each replica are determined by the linear relation, $(T_a - T) / (T_{DEC} - T) = (\lambda_a - 1) / (\lambda_{DEC} - 1) = (a - 1) / (N_{\mathrm{REP}} - 1)$, with $a = 1, 2, \ldots, N_{\mathrm{REP}}$. Note that the overlap function, $Q_\alpha(t,d)$ [Eq. (13)], between two structures is evaluated using only the structures generated in the production sweeps of the replica with $a = 1$, i.e., $T_1 = T$ and $\lambda_1 = 1$, where $t$ is replaced by the number of production sweeps separating them. See ref [1] for details. The same definitions also apply to Tables S4–S8.

Table S4. Replica parameters for $T = 0.698$.

| $d$ | 1.5 | 1.7 | 1.9 | 2.1 | 2.3 | 2.5 | 2.7 | 2.9 | 3.1 | 3.3 |
|---|---|---|---|---|---|---|---|---|---|---|
| $S_{EQ}$ | 1000 | 1000 | 1000 | 1000 | 1000 | 1000 | 3000 | 3000 | 3000 | 3000 |
| $S_{PROD}$ | 4000 | 4000 | 4000 | 4000 | 7000 | 9000 | 7000 | 7000 | 13000 | 13000 |
| $T_{DEC}$ | 1.0 | 1.0 | 1.0 | 1.0 | 0.8 | 0.8 | 0.8 | 0.8 | 0.8 | 0.8 |
| $\lambda_{DEC}$ | 0.80 | 0.80 | 0.80 | 0.80 | 0.82 | 0.86 | 0.86 | 0.88 | 0.88 | 0.89 |
| $N_{REP}$ | 8 | 11 | 11 | 11 | 16 | 16 | 16 | 16 | 19 | 19 |

Table S5. Replica parameters for $T = 0.550$.

| $d$ | 1.5 | 1.7 | 1.9 | 2.1 | 2.3 | 2.5 | 2.7 | 2.9 | 3.1 | 3.3 | 3.5 |
|---|---|---|---|---|---|---|---|---|---|---|---|
| $S_{EQ}$ | 1000 | 1000 | 1000 | 1000 | 1000 | 1000 | 3000 | 3000 | 3000 | 4000 | 4000 |
| $S_{PROD}$ | 4000 | 4000 | 4000 | 4000 | 7000 | 9000 | 7000 | 7000 | 13000 | 16000 | 16000 |
| $T_{DEC}$ | 1.0 | 1.0 | 1.0 | 1.0 | 0.8 | 0.8 | 0.8 | 0.8 | 0.6 | 0.6 | 0.6 |
| $\lambda_{DEC}$ | 0.80 | 0.80 | 0.80 | 0.80 | 0.82 | 0.86 | 0.86 | 0.88 | 0.88 | 0.89 | 0.90 |
| $N_{REP}$ | 8 | 11 | 11 | 11 | 16 | 16 | 16 | 16 | 19 | 19 | 19 |

Table S6. Replica parameters for $T = 0.511$. $d = 3.7$ and $3.9$ were computed using MD.

| $d$ | 1.5 | 1.7 | 1.9 | 2.1 | 2.3 | 2.5 | 2.7 | 2.9 | 3.1 | 3.3 | 3.5 |
|---|---|---|---|---|---|---|---|---|---|---|---|
| $S_{EQ}$ | 1000 | 1000 | 1000 | 1000 | 1000 | 1000 | 1000 | 3000 | 3000 | 3000 | 4000 |
| $S_{PROD}$ | 4000 | 4000 | 4000 | 4000 | 7000 | 9000 | 9000 | 7000 | 13000 | 13000 | 16000 |
| $T_{DEC}$ | 1.0 | 1.0 | 1.0 | 1.0 | 0.8 | 0.8 | 0.8 | 0.8 | 0.8 | 0.6 | 0.6 |
| $\lambda_{DEC}$ | 0.80 | 0.80 | 0.80 | 0.80 | 0.82 | 0.86 | 0.86 | 0.88 | 0.88 | 0.89 | 0.90 |
| $N_{REP}$ | 8 | 11 | 11 | 13 | 16 | 16 | 16 | 16 | 19 | 19 | 19 |

Table S7. Replica parameters for $T = 0.476$.

| $d$ | 1.5 | 1.7 | 1.9 | 2.1 | 2.3 | 2.5 | 2.7 | 2.9 | 3.1 | 3.3 | 3.5 | 3.7 |
|---|---|---|---|---|---|---|---|---|---|---|---|---|
| $S_{EQ}$ | 1000 | 1000 | 1000 | 1000 | 1000 | 2000 | 2000 | 3000 | 3000 | 3000 | 4000 | 4000 |
| $S_{PROD}$ | 4000 | 4000 | 4000 | 4000 | 4000 | 6000 | 6000 | 13000 | 13000 | 17000 | 16000 | 16000 |
| $T_{DEC}$ | 1.0 | 1.0 | 1.0 | 1.0 | 0.8 | 0.8 | 0.8 | 0.6 | 0.6 | 0.6 | 0.6 | 0.6 |
| $\lambda_{DEC}$ | 0.80 | 0.80 | 0.80 | 0.80 | 0.82 | 0.86 | 0.86 | 0.88 | 0.88 | 0.89 | 0.90 | 0.90 |
| $N_{REP}$ | 8 | 11 | 11 | 13 | 16 | 16 | 16 | 19 | 19 | 19 | 19 | 19 |

Table S8. Replica parameters for $T = 0.455$. $d = 3.9$ was computed using MD.

| $d$ | 1.5 | 1.7 | 1.9 | 2.1 | 2.3 | 2.5 | 2.7 | 2.9 | 3.1 | 3.3 | 3.5 | 3.7 |
|---|---|---|---|---|---|---|---|---|---|---|---|---|
| $S_{EQ}$ | 1000 | 1000 | 1000 | 1000 | 1000 | 2000 | 2000 | 3000 | 3000 | 3000 | 4000 | 4000 |
| $S_{PROD}$ | 4000 | 4000 | 4000 | 4000 | 4000 | 6000 | 6000 | 13000 | 13000 | 17000 | 16000 | 16000 |
| $T_{DEC}$ | 1.0 | 1.0 | 1.0 | 1.0 | 0.8 | 0.8 | 0.8 | 0.6 | 0.6 | 0.6 | 0.6 | 0.6 |
| $\lambda_{DEC}$ | 0.80 | 0.80 | 0.80 | 0.80 | 0.82 | 0.86 | 0.86 | 0.88 | 0.88 | 0.89 | 0.90 | 0.90 |
| $N_{REP}$ | 8 | 11 | 11 | 13 | 16 | 16 | 16 | 19 | 19 | 19 | 19 | 19 |

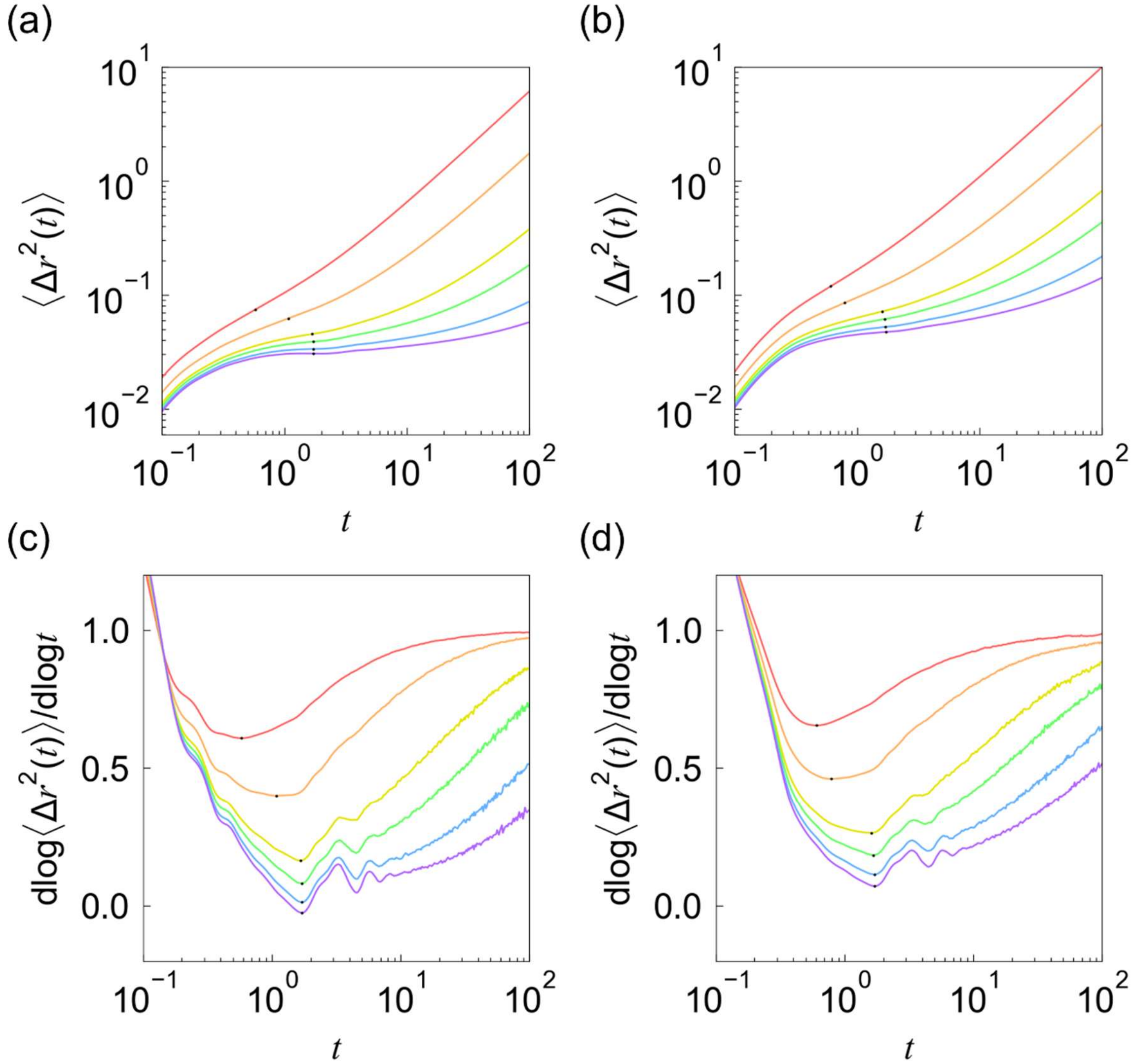

Figure S1. (a) and (b) Mean-squared displacements (MSDs) for particles A and B, respectively. (c) and (d) Derivatives of the logarithmic MSDs with respect to logarithmic time for particles A and B, respectively. In (a)-(d), colors represent the results at $T = 0.982$ (red), 0.698 (orange), 0.550 (yellow), 0.511 (green), 0.476 (blue), and 0.455 (purple), respectively. Black dots indicate the $\Delta t$ values used in the definition of the hop function.

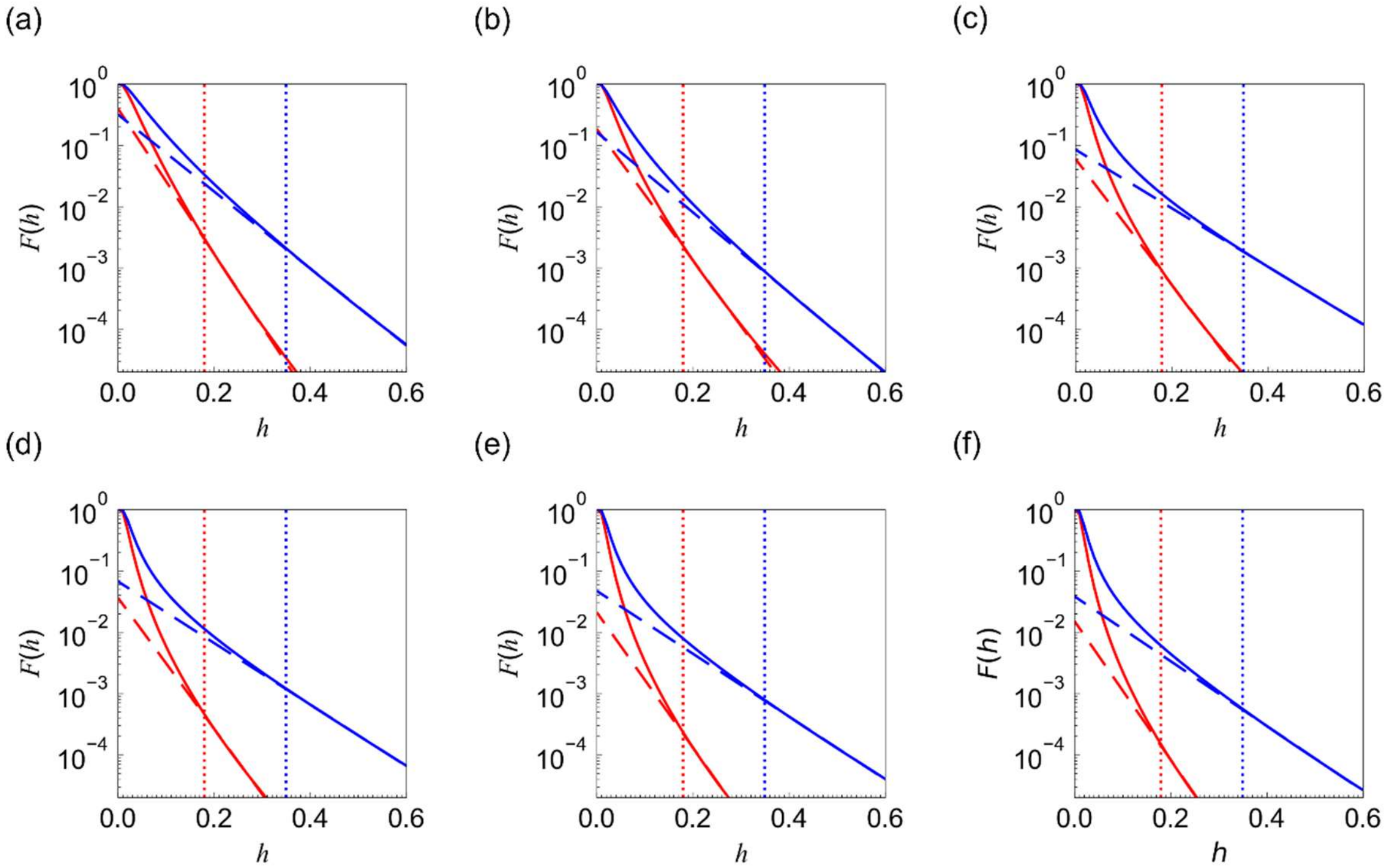

Figure S2. Cumulative probability of $h$, $F(h)$ (solid), of particles A (red) and B (blue), and exponential fit to the tail (dashed) at $T$ = (a) 0.982, (b) 0.698, (c) 0.550, (d) 0.511, (e) 0.476, and (f) 0.455. Vertical dotted lines indicate the threshold values of $h^*$.

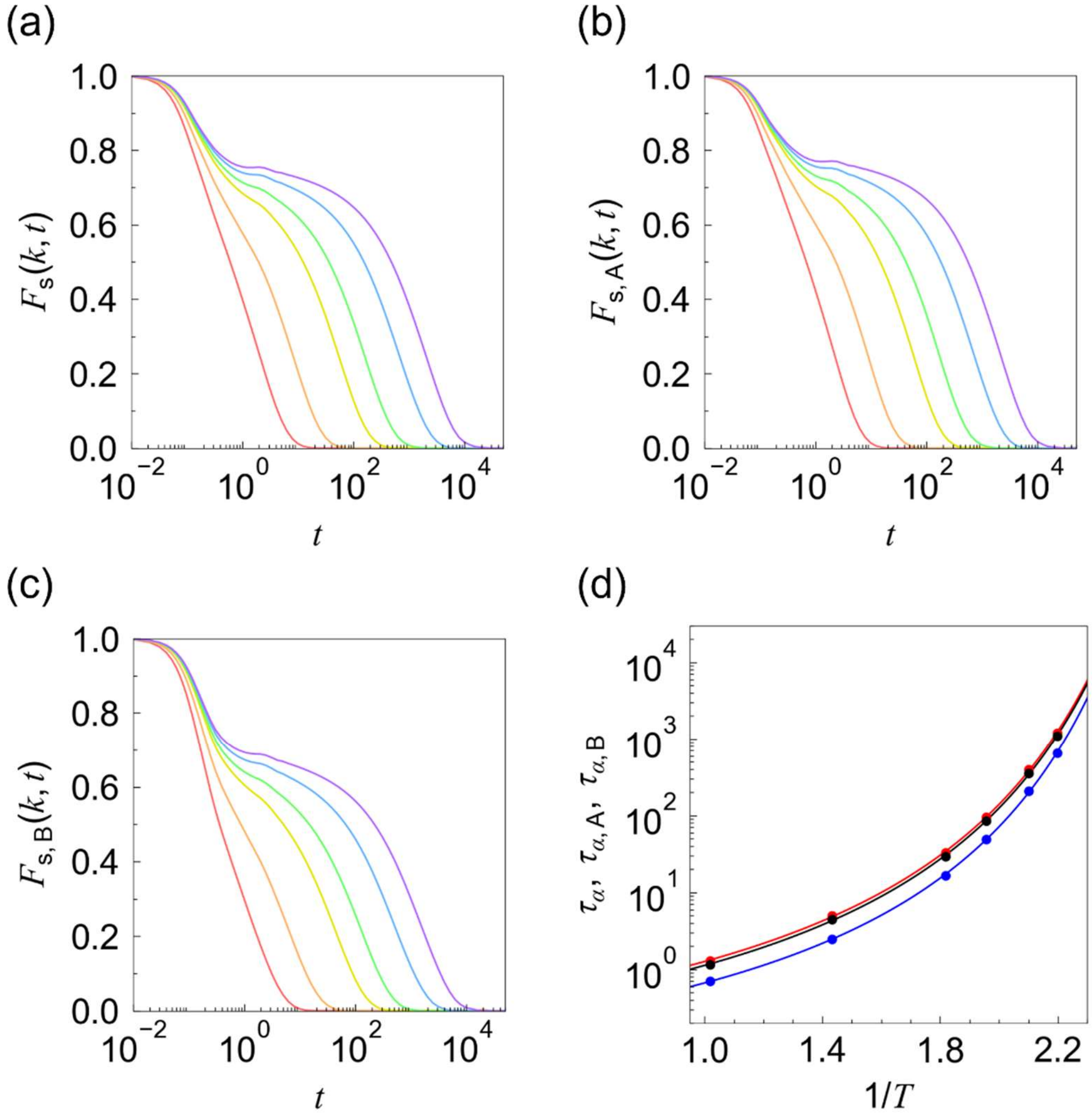

Figure S3. (a)-(c) Self-intermediate scattering functions $F_s(k, t)$ for both species, $F_{s,A}(k, t)$ for particle A, and $F_{s,B}(k, t)$ for particle B, respectively. (d) Relaxation times, $\tau_\alpha$, for particle A (red), particle B (blue), and both species (black). In (d), the curves are fitted using the Vogel–Fulcher–Tammann (VFT) equation,[2-4] and the obtained kinetic fragilities for particle A (red), particle B (blue), and both species (black) are 0.290, 0.289, and 0.307, respectively.

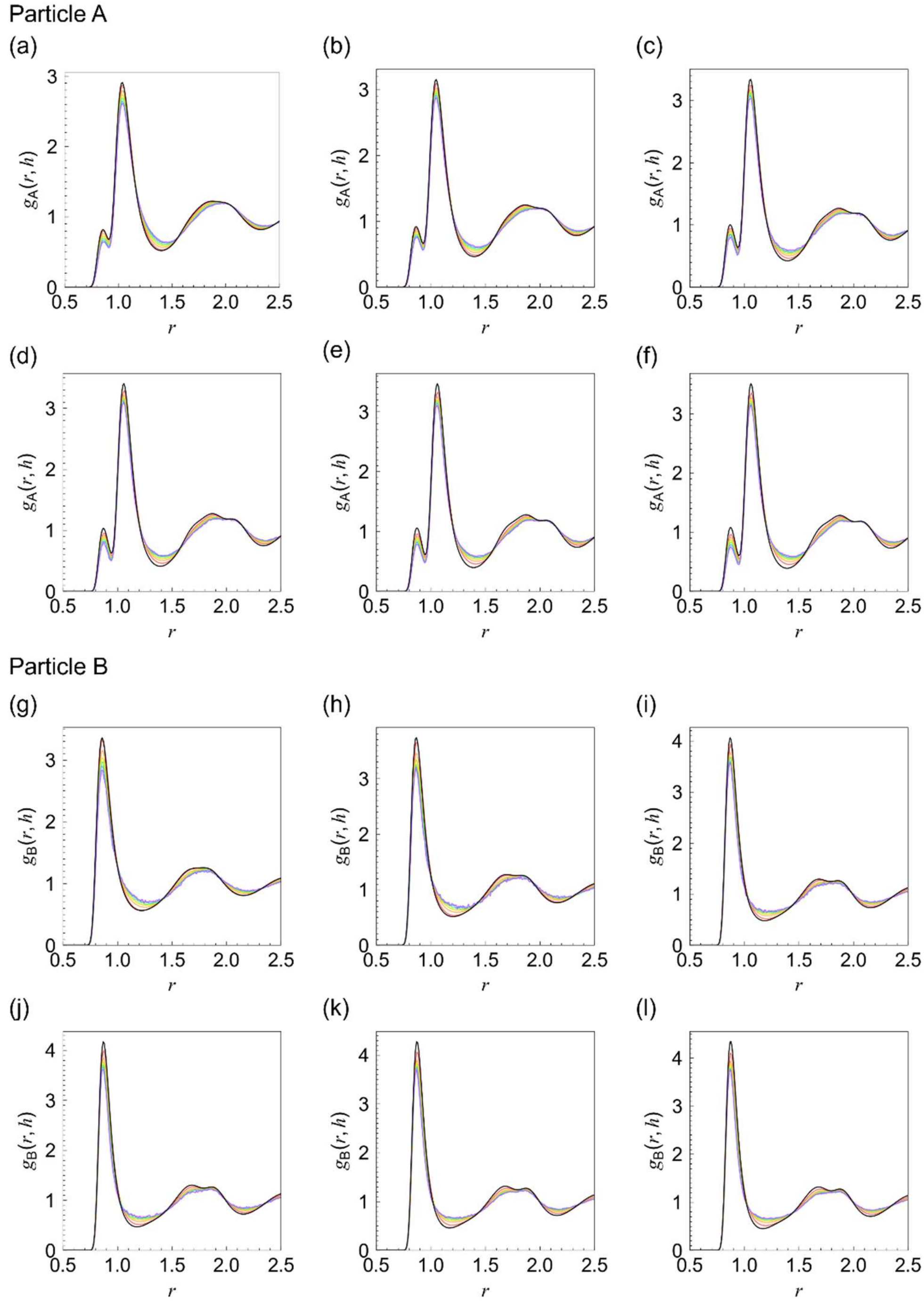

Figure S4. $h$-dependent radial distribution function of particle A, $g_A(r,h)$, at $T$ = 0.982 (a), 0.698 (b), 0.550 (c), 0.511 (d), 0.476 (e), and 0.455 (f), and of particle B, $g_B(r,h)$, at $T$ = 0.982 (g), 0.698 (h), 0.550 (i), 0.511 (j), 0.476 (k), and 0.455 (l). See Fig. 1 (b) and (f) for the color codes.

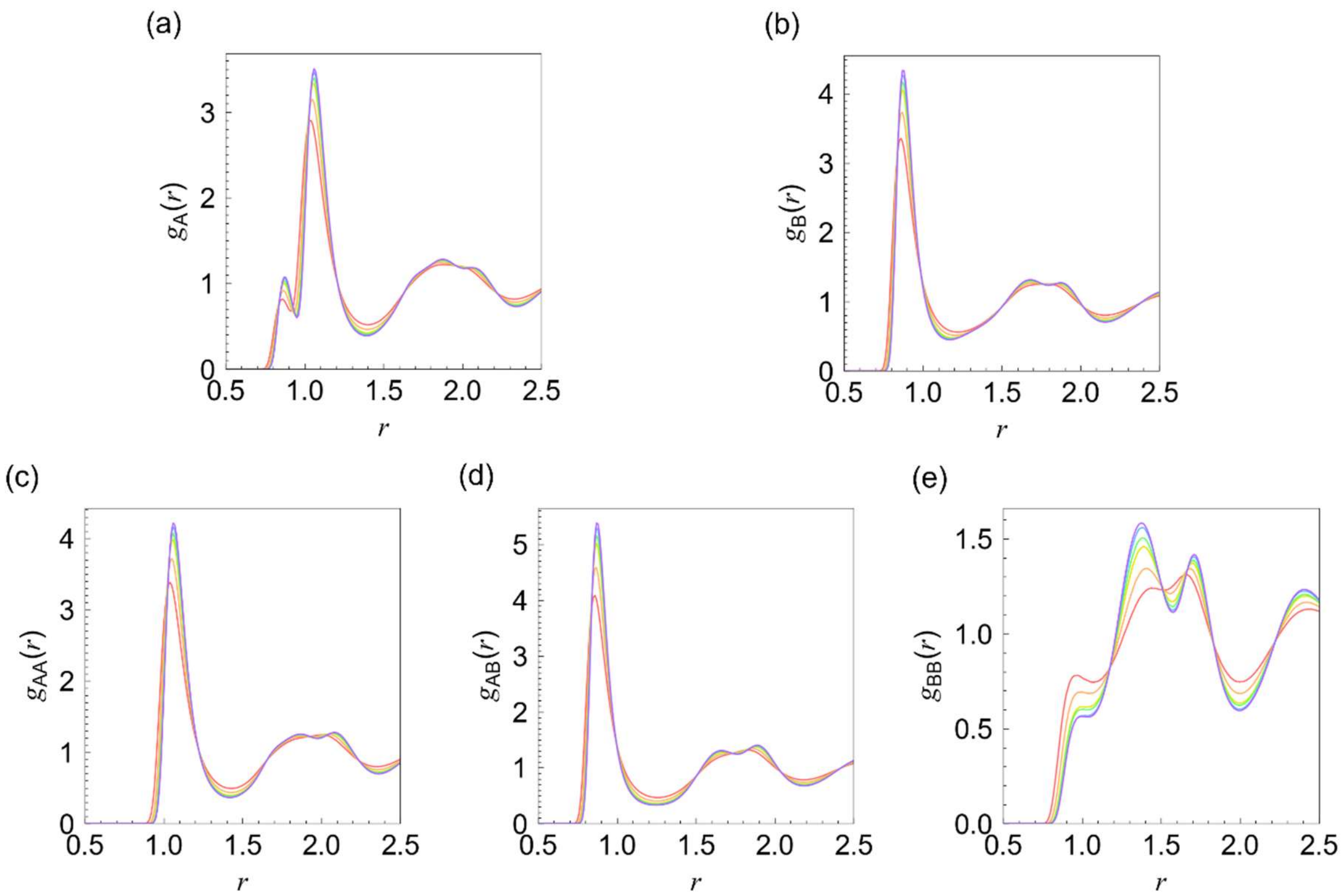

Figure S5. Radial distribution functions, (a) $g_A(r)$, (b) $g_B(r)$, (c) $g_{AA}(r)$, (d) $g_{AB}(r)$, and (e) $g_{BB}(r)$ at $T$ = 0.982 (red), 0.698 (orange), 0.550 (yellow), 0.511 (green), 0.476 (blue), and 0.455 (purple).

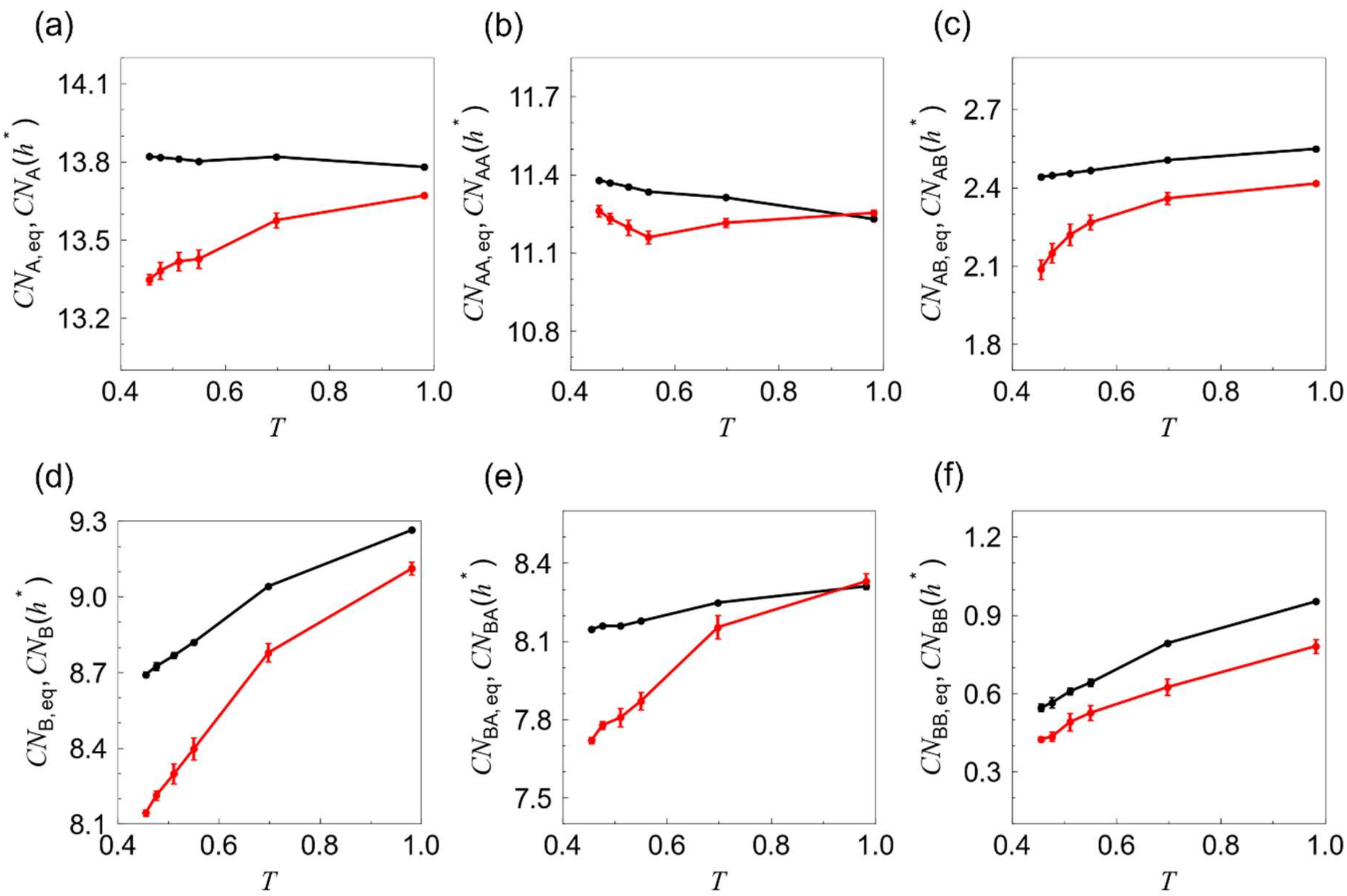

Figure S6. Temperature-dependent coordination numbers at equilibrium (black) and at $h^*$ (red). (a) $CN_{A,eq}$ and $CN_A(h^*)$, (b) $CN_{AA,eq}$ and $CN_{AA}(h^*)$, (c) $CN_{AB,eq}$ and $CN_{AB}(h^*)$, (d) $CN_{B,eq}$ and $CN_B(h^*)$, (e) $CN_{BA,eq}$ and $CN_{BA}(h^*)$, and (f) $CN_{BB,eq}$ and $CN_{BB}(h^*)$. The error bars were evaluated using the bootstrap method.

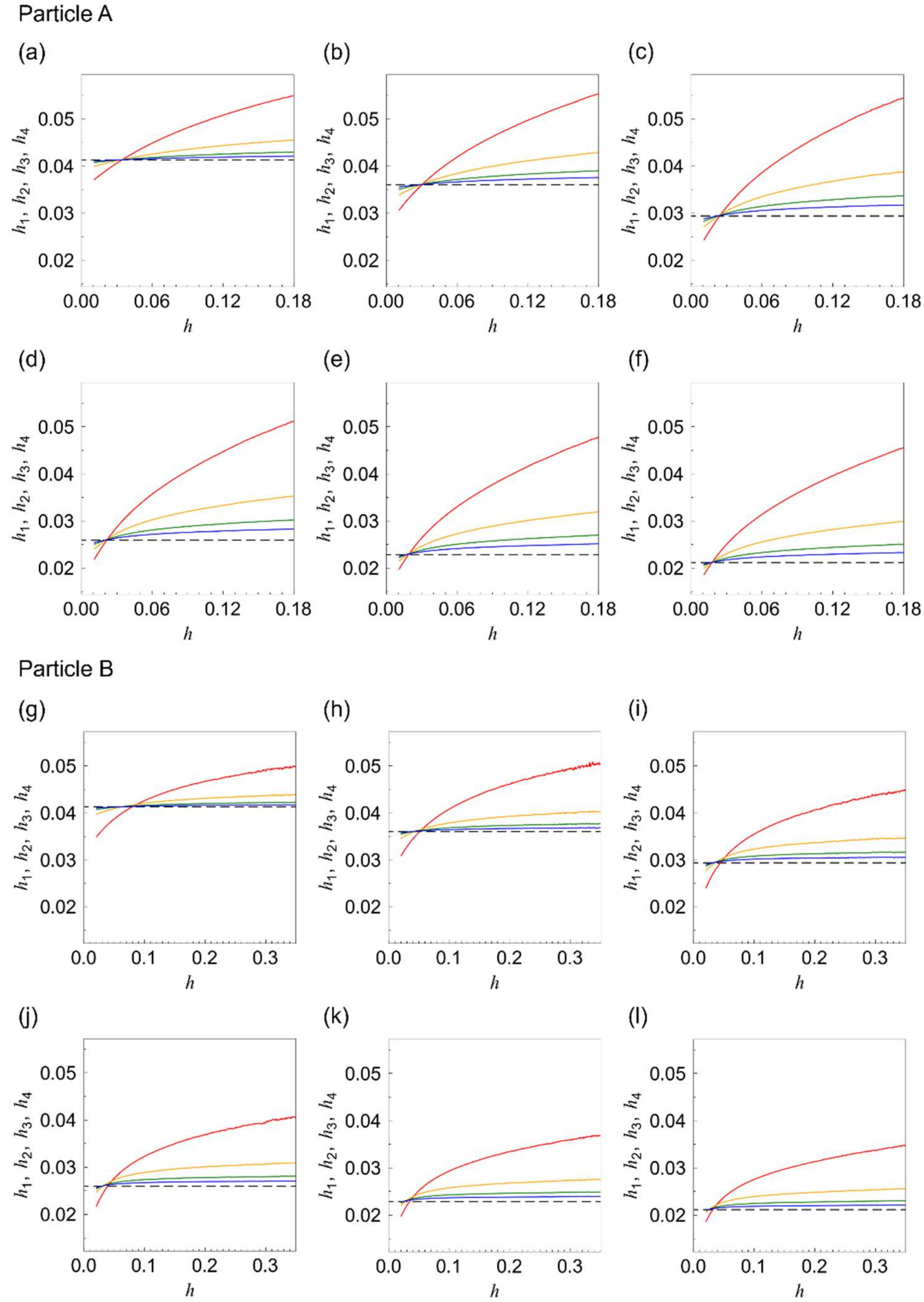

Figure S7. $h$ values averaged over neighboring particles within the $n$-th coordination shell ($n$ = 1 - 4) plotted against the $h$ value of the jumping particle A at (a) $T$ = 0.982, (b) $T$ = 0.698, (c) $T$ = 0.550, (d) $T$ = 0.511, (e) $T$ = 0.476, and (f) $T$ = 0.455, and of the jumping particle B at (g) $T$ = 0.982, (h) $T$ = 0.698, (i) $T$ = 0.550, (j) $T$ = 0.511, (k) $T$ = 0.476, and (l) $T$ = 0.455. Red, orange, green, and blue indicate the first to the fourth coordination shells. Black dashed lines indicate $h_{eq}$.

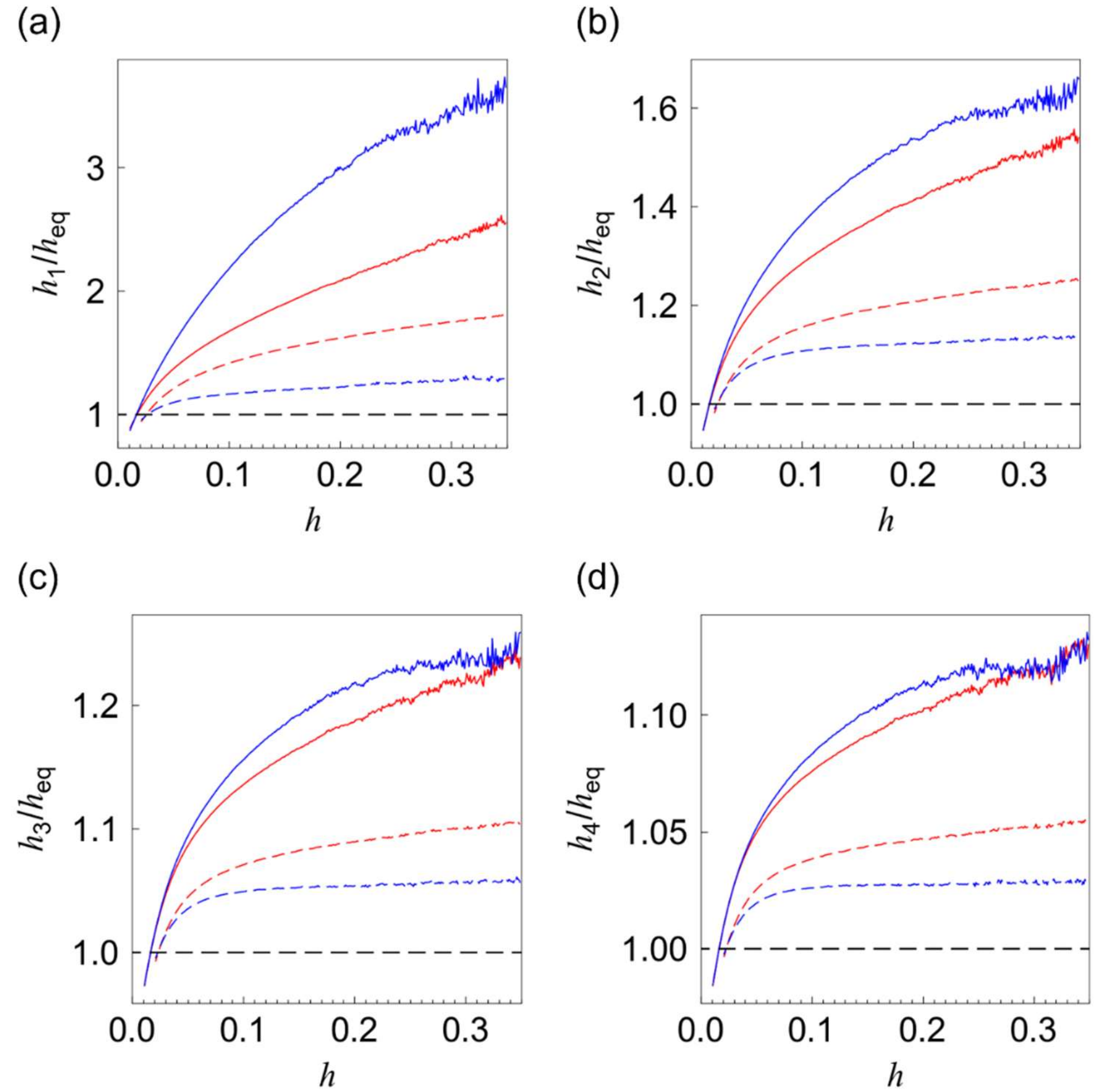

Figure S8. $h$ values of neighboring particles A and B plotted against the $h$ values of the jumping particles A and B at T = 0.455. Solid red and solid blue curves represent averages around jumping particle A over neighboring particles A and B, respectively; dashed red and dashed blue curves represent averages around jumping particle B over neighboring particles A and B, respectively. Panels (a)–(d) correspond to the first through fourth coordination shells.

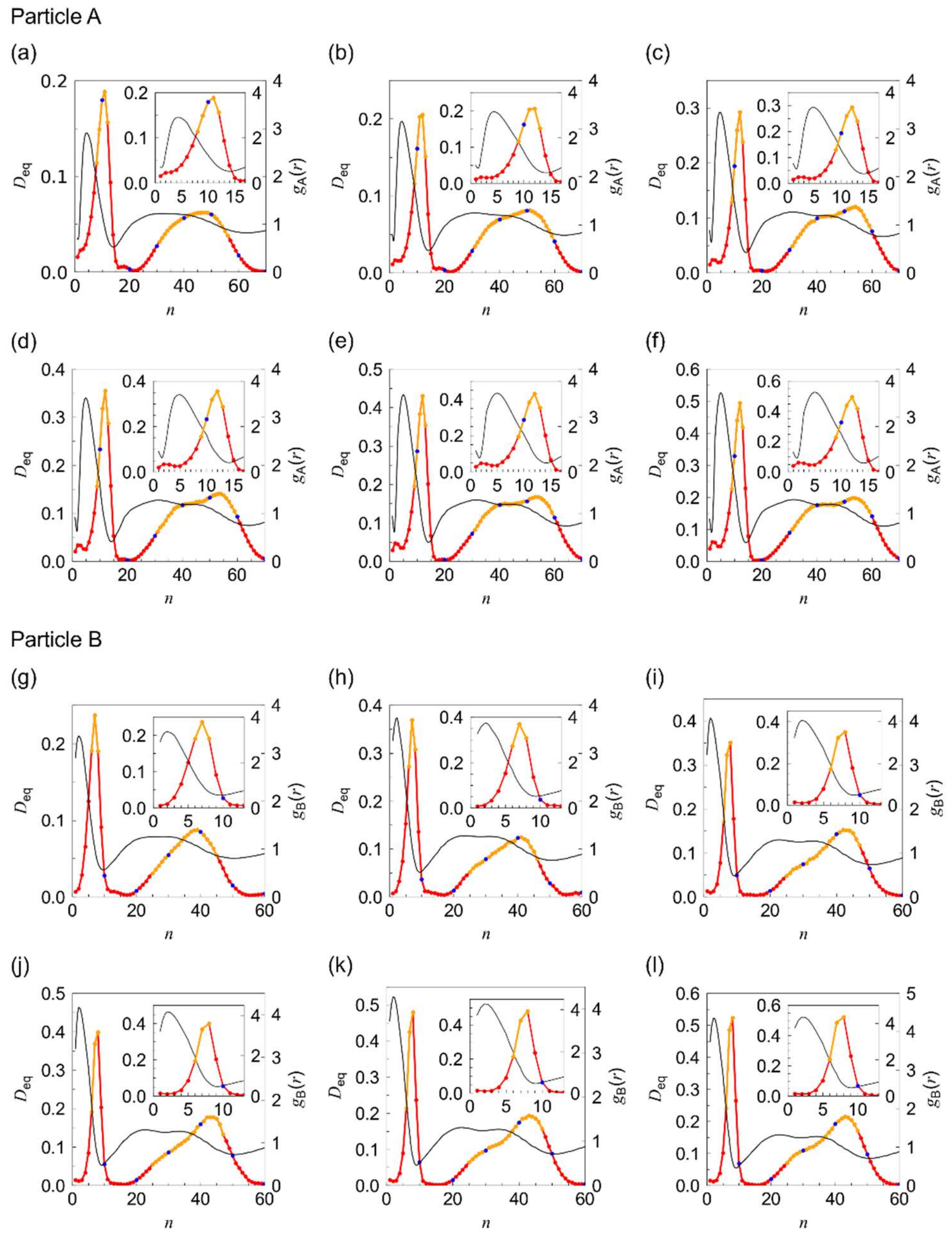

Figure S9. $D_{\text{eq}}$ (red) and $g(f(\langle r_n \rangle))$ (black) of neighboring particle $n$, plotted against $n$, where $\langle r_n \rangle = \int r_n P_{\text{eq}}(r_n) \mathrm{d}r_n$ and $f(\langle r_n \rangle)$ is a piece-wise linear function that maps $\langle r_n \rangle$ to $n$. Panels (a-f) correspond to particle A at (a) $T = 0.982$, (b) $T = 0.698$, (c) $T = 0.550$, (d) $T = 0.511$, (e) $T = 0.476$, and (f) $T = 0.455$. Panels (g-l) show the same quantities for particle B at the corresponding temperatures. Blue dots mark every tenth particle. Yellow dots mark the particles involved in the collective variables *CV1* and *CV2*. The insets show zoomed-in views for the first peaks. Note that (a), (f), (g), and (l) are the same as Figs. 6 (b), 6 (a), 6 (d), and 6 (c), respectively.

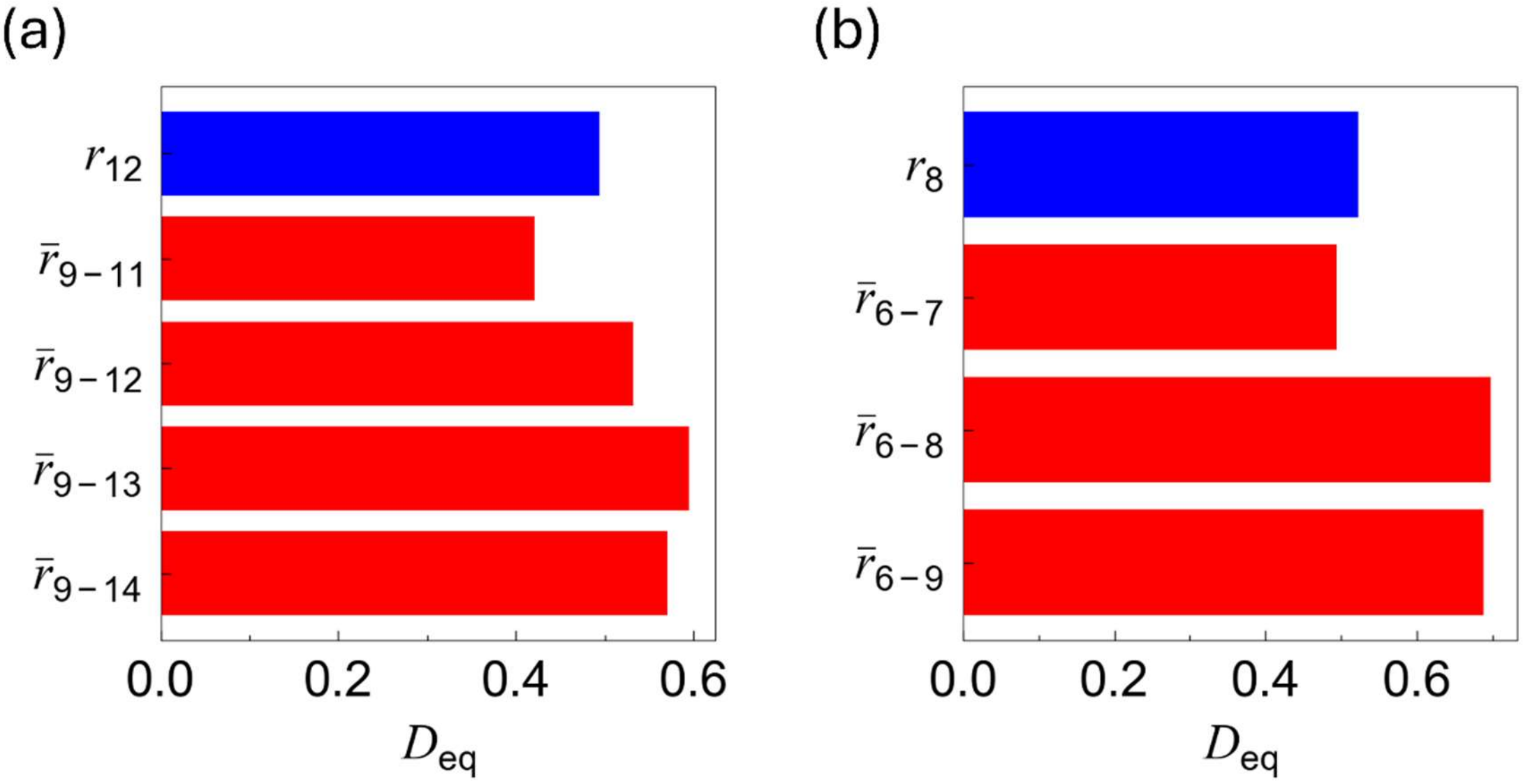

Figure S10. $D_{eq}$ values of a single particle (blue) and the values of important collective variables (red) for (a) particle A and (b) particle B at $T = 0.455$.

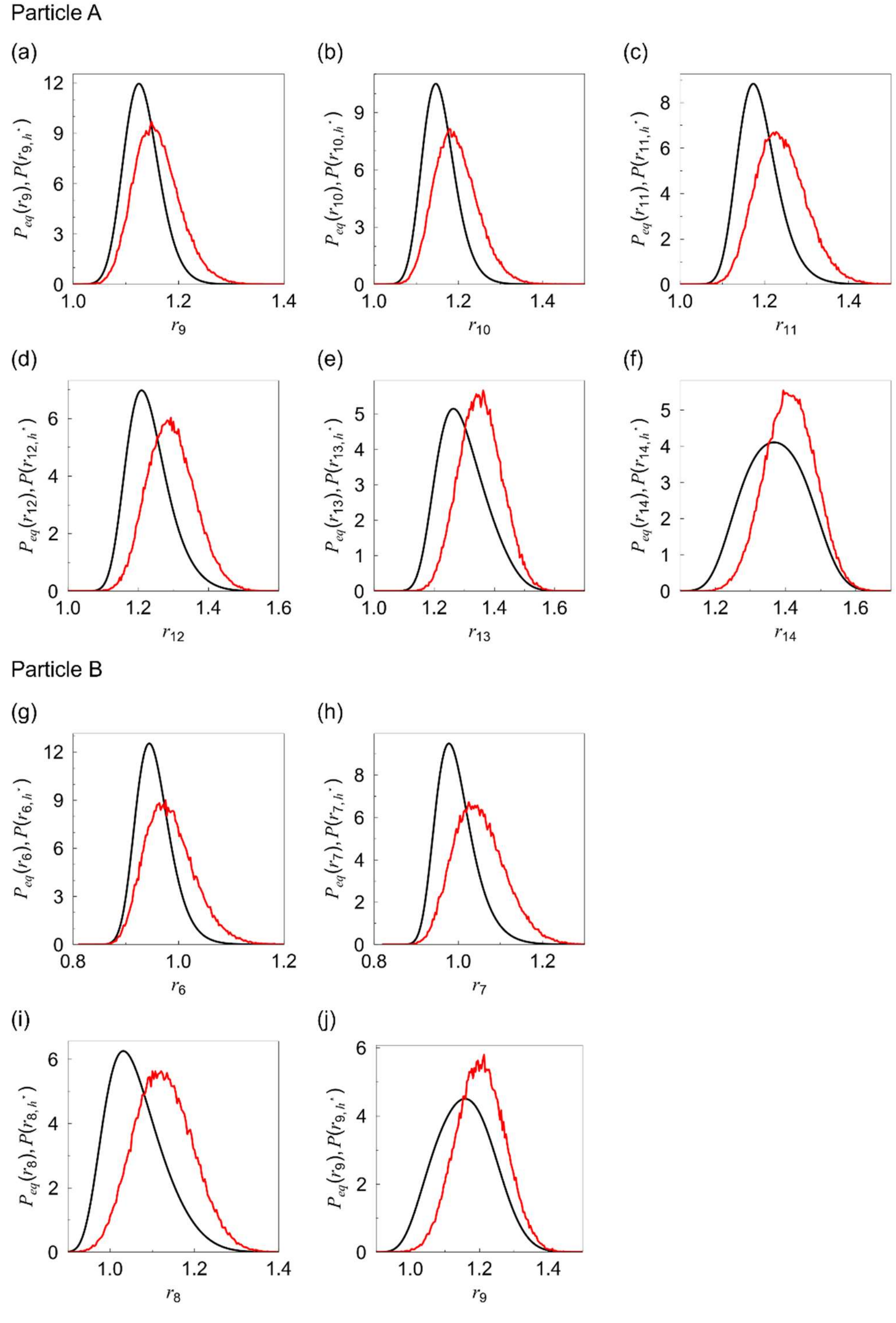

Figure S11. Distributions $P_{\mathrm{eq}}(r_n)$ (black) and $P(r_n,h^*)$ (red) for the ninth to 14th nearest neighbors to a jumping particle A (a-f) and for the sixth to ninth nearest neighbors to a jumping particle B (g-j) at $T = 0.455$.

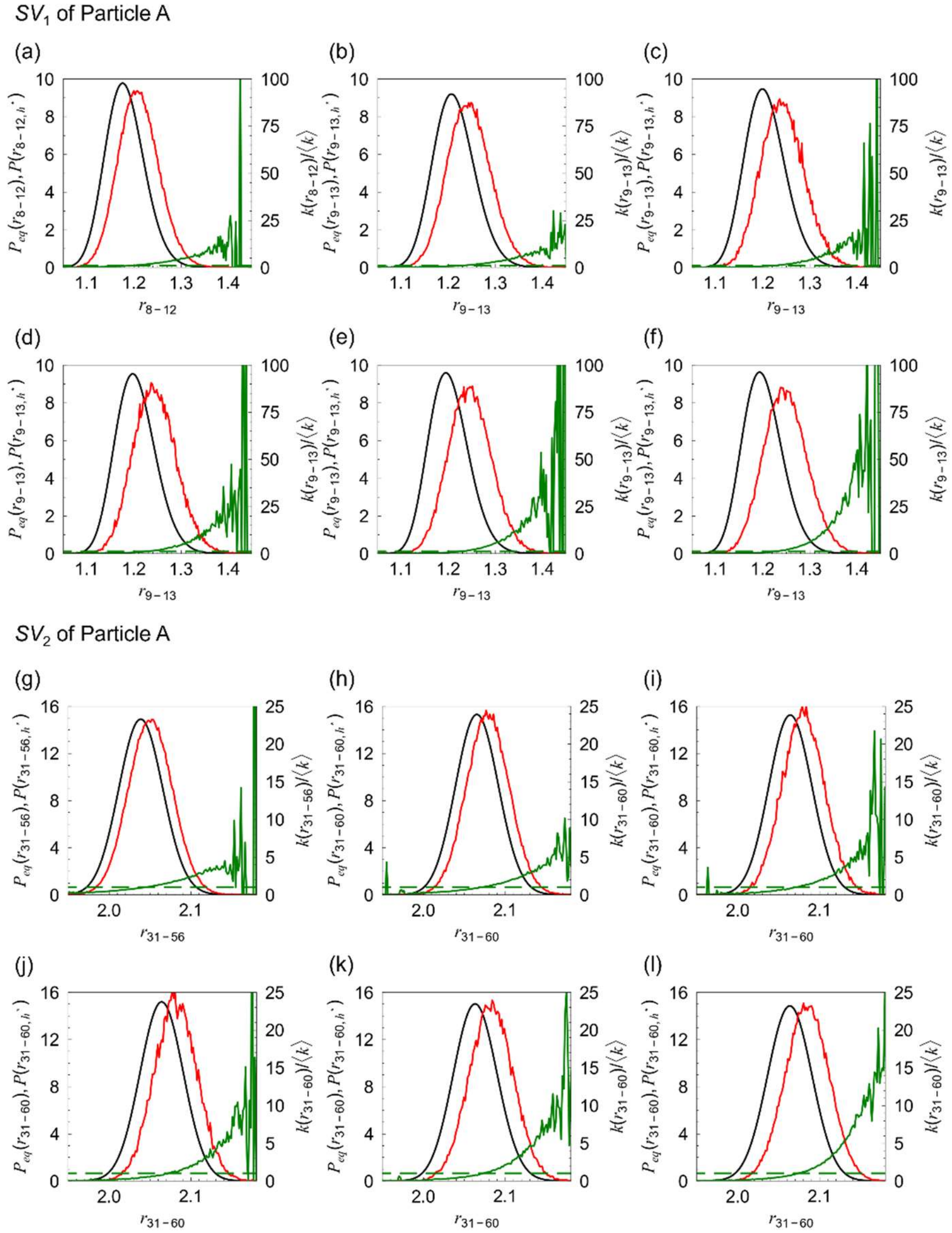

Figure S12. Distributions $P_{\mathrm{eq}}(r_n)$ (black) and $P(r_n,h^*)$ (red) for the average distance of neighbors used in *CV1* and *CV2* of particle A at equilibrium and at $h^*$, respectively. The relative rate $k(r_n)/\langle k\rangle$ (green) obtained from these distributions using $k_i = N_i^*/(N_i\delta t)$. Panels (a-f) are for *CV1*. (a) $T = 0.982$, $n = 8{-}12$. (b) $T = 0.698$, $n = 9{-}13$. (c) $T = 0.550$, $n = 9{-}13$. (d) $T = 0.511$, $n = 9{-}13$. (e) $T = 0.476$, $n = 9{-}13$. (f) $T = 0.455$, $n = 9{-}13$. Panels (g-l) are for *CV2*. (g) $T = 0.982$, $n = 31{-}56$. (h) $T = 0.698$, $n = 31{-}60$. (i) $T = 0.550$, $n = 31{-}60$. (j) $T = 0.511$, $n = 31{-}60$. (k) $T = 0.476$, $n = 31{-}60$. (l) $T = 0.455$, $n = 31{-}60$. The green dashed line indicates $\langle k\rangle/\langle k\rangle = 1$. The $D_{\mathrm{eq}}$ values for *CV1* at these temperatures are 0.238, 0.258, 0.349, 0.429, 0.513, and 0.595, respectively, whereas those for *CV2* are 0.094, 0.123, 0.184, 0.226, 0.271, and 0.309.

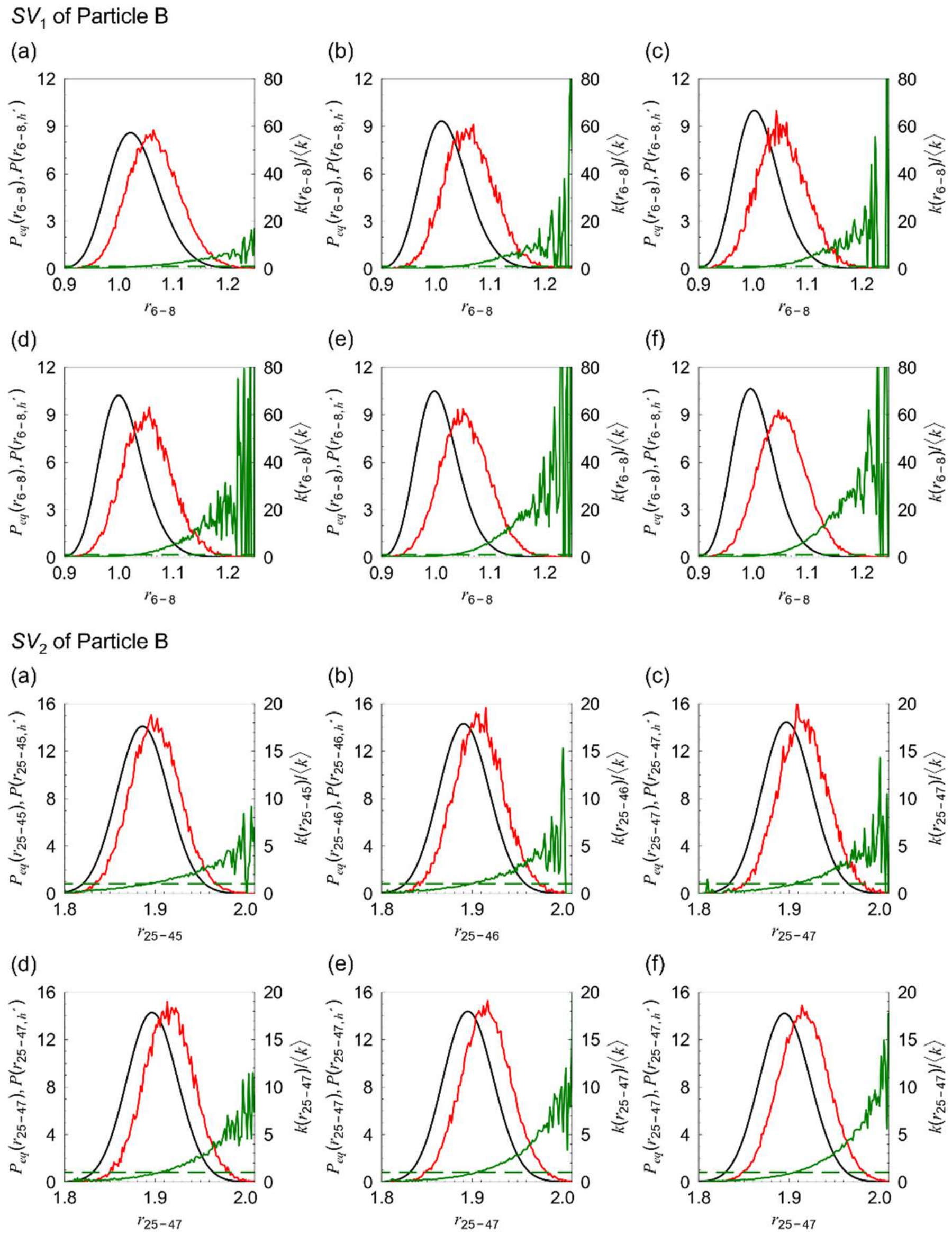

Figure S13. Distributions $P_{\mathrm{eq}}(r_n)$ (black) and $P(r_n,h^*)$ (red) for the average distance of neighbors used in *CV1* and *CV2* of particle B at equilibrium and at $h^*$, respectively. The relative rate $k(r_n)/\langle k\rangle$ (green) obtained from these distributions using $k_i = N_i^*/(N_i\delta t)$. Panels (a-f) are for *CV1*. (a) $T = 0.982$, $n = 6-8$. (b) $T = 0.698$, $n = 6-8$. (c) $T = 0.550$, $n = 6-8$. (d) $T = 0.511$, $n = 6-8$. (e) $T = 0.476$, $n = 6-8$. (f) $T = 0.455$, $n = 6-8$. Panels (g-l) are for *CV2*. (g) $T = 0.982$, $n = 25-45$. (h) $T = 0.698$, $n = 25-46$. (i) $T = 0.550$, $n = 25-46$. (j) $T = 0.511$, $n = 25-47$. (k) $T = 0.476$, $n = 25-47$. (l) $T = 0.455$, $n = 25-47$. The green dashed line indicates $\langle k\rangle/\langle k\rangle = 1$. The $D_{\mathrm{eq}}$ values for *CV1* at these temperatures are 0.301, 0.485, 0.453, 0.537, 0.626, and 0.698, respectively, whereas those for *CV2* are 0.118, 0.166, 0.193, 0.222, 0.248, and 0.264.

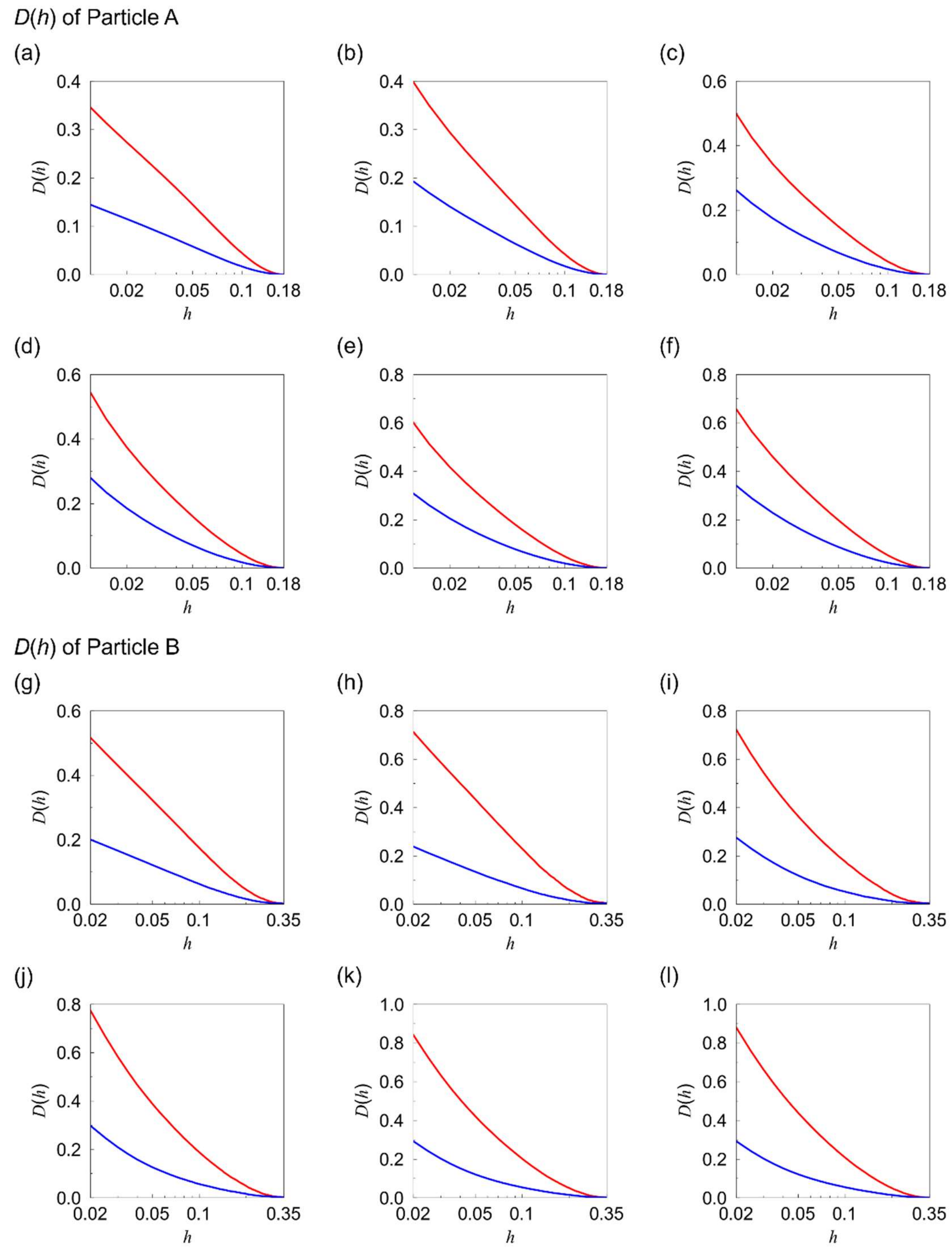

Figure S14. KL divergence $D(h)$ of collective variables *CV1* (red) and *CV2* (blue) for particle A (a-f) and particle B (g-l) at (a) and (g) $T = 0.982$, (b) and (h) $T = 0.698$, (c) and (i) $T = 0.550$, (d) and (j) $T = 0.511$, (e) and (k) $T = 0.476$, and (f) and (l) $T = 0.455$.

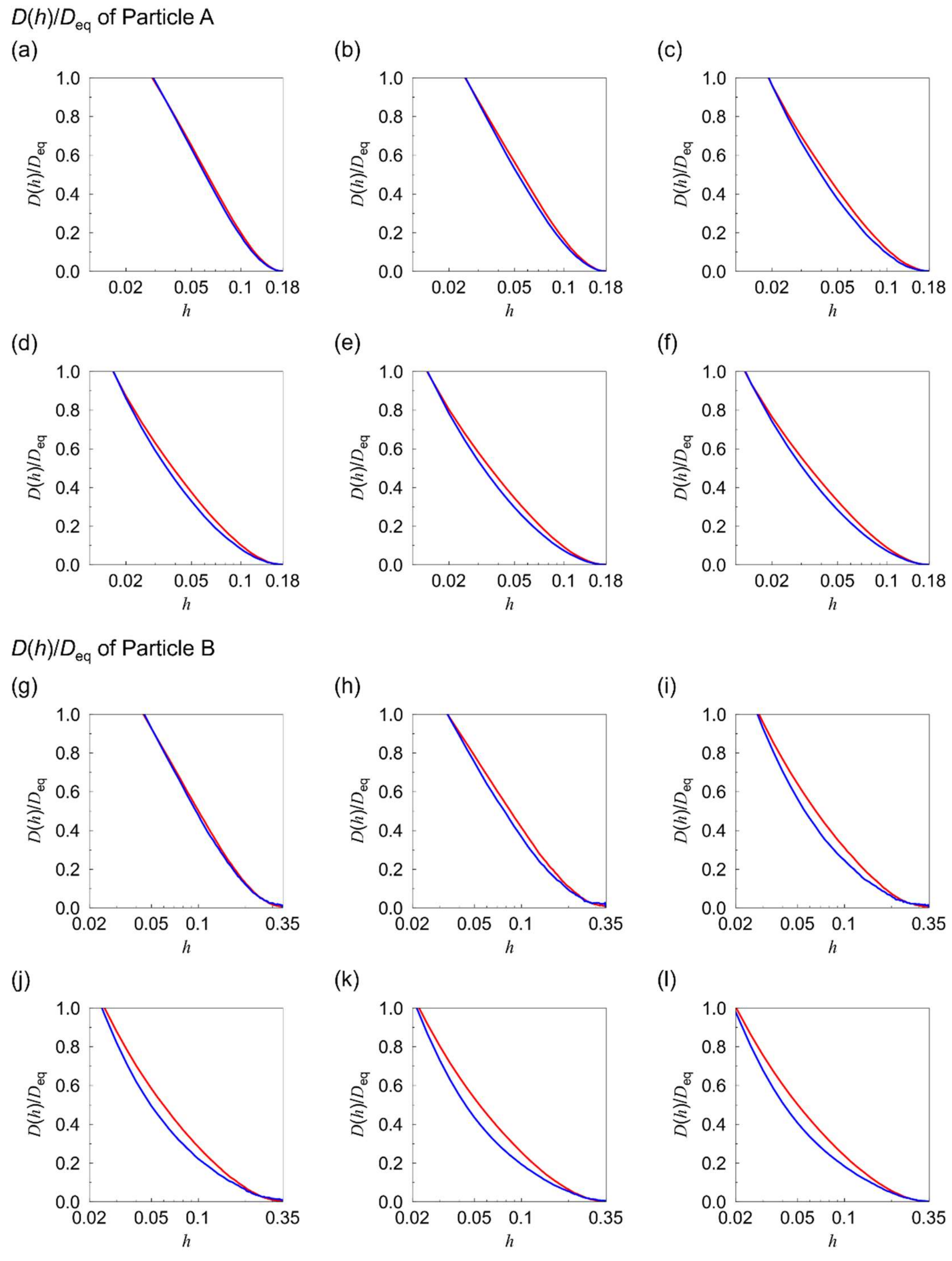

Figure S15. KL divergence $D(h)$ normalized by $D_{eq}$ of collective variables *CV1* (red) and *CV2* (blue) for particle A (a-f) and particle B (g-l) at (a) and (g) $T = 0.982$, (b) and (h) $T = 0.698$, (c) and (i) $T = 0.550$, (d) and (j) $T = 0.511$, (e) and (k) $T = 0.476$, and (f) and (l) $T = 0.455$.

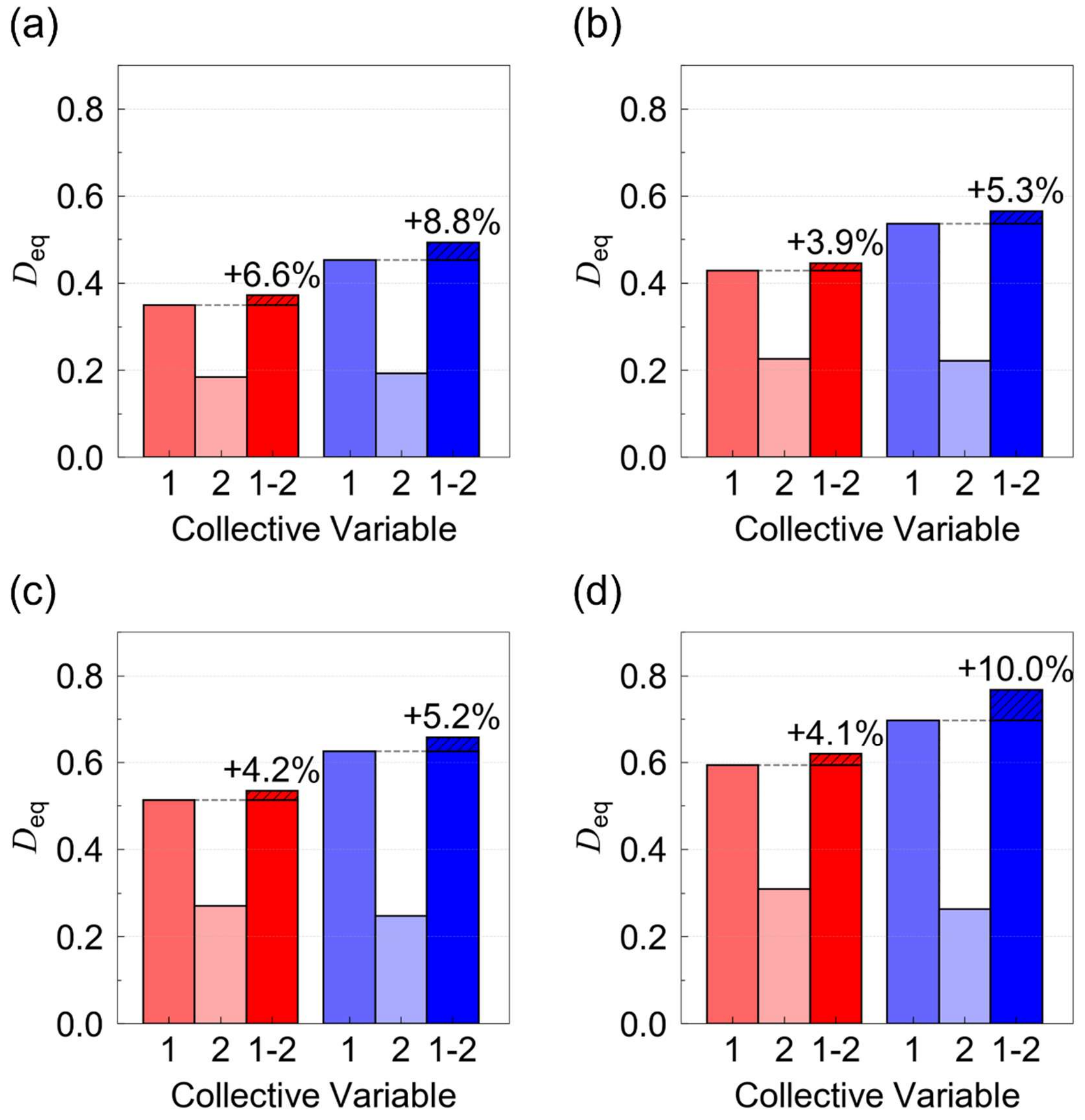

Figure S16. Comparison of $D_{\mathrm{eq}}$ for the collective variables, *CV1*, *CV2*, and *CV1-2* at (a) $T = 0.550$, (b) $T = 0.511$, (c) $T = 0.476$, and (d) $T = 0.455$ for particles A (red) and B (blue). (d) is identical to Fig. 10.

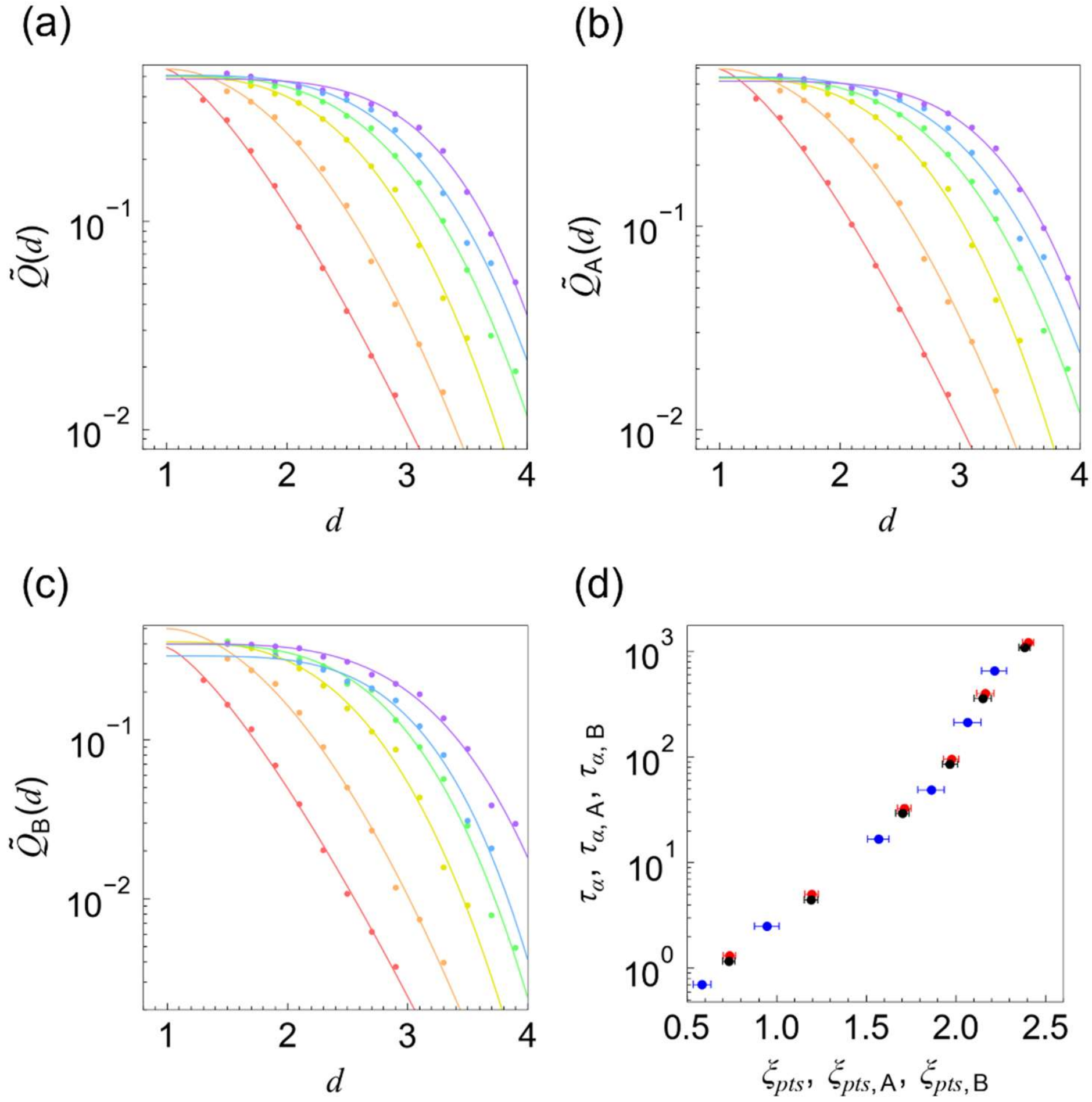

Figure S17. (a)-(c) $\widetilde{Q}(d)$, $\widetilde{Q}_{\mathrm{A}}(d)$, and $\widetilde{Q}_{\mathrm{B}}(d)$, respectively, at the six temperatures. (d) Temperature dependence of $\xi_{\mathrm{pts}}$ against $\tau_{\alpha}$ for central particle A (red), particle B (blue), and both species without distinguishing between particles A and B (black). In (a)-(c), the solid curves represent fits using Eq. (14), from which $\xi_{\mathrm{pts},\alpha}$ values were obtained. The color codes are the same as those in Fig. 2. In (d), the standard deviations were calculated using the bootstrap method.

## References

1. L. Berthier, P. Charbonneau and S. Yaida, J. Chem. Phys. **144**, 024501 (2016).
2. H. Vogel, Phys. Z. **22** (1921).
3. G. S. Fulcher, J. Am. Ceram. Soc. **8** (1925).
4. G. Tammann and W. Hesse, Z. Anorg. Allg. Chem. **156** (1926).